\documentclass[12pt]{article}

\usepackage{amsmath}
\usepackage{graphicx}
\usepackage{amsfonts}
\usepackage{amssymb}
\usepackage{epsfig}
\usepackage{color}
\usepackage{psfrag}

\newcommand{\EQ}{\begin{equation}}
\newcommand{\EN}{\end{equation}}

\def\th{\theta}

\begin{document}
\setcounter{page}{0} \topmargin 0pt
\renewcommand{\thefootnote}{\arabic{footnote}}
\newpage
\setcounter{page}{0}

\begin{titlepage}

\begin{flushright}
ISAS/1/2004/FM
\end{flushright}
\vspace{0.5cm}
\begin{center}
{\Large {\bf Semiclassical Particle Spectrum }}\\
{\Large {\bf of Double Sine--Gordon Model}} \\

\vspace{15mm} {\large G. Mussardo$^{1,2}$, V. Riva$^{1,2}$
 and G. Sotkov$^{3}$} \\
\vspace{0.5cm} {\em $^{1}$International School for Advanced Studies}\\
{\em Via Beirut 1, 34100 Trieste, Italy} \\
\vspace{0.3cm} {\em $^{2}$Istituto Nazionale di Fisica Nucleare}\\
{\em Sezione di Trieste}\\\vspace{0.3cm} {\em $^3$ Instituto de Fisica Teorica} \\
{\em Universidade Estadual Paulista} \\
{\em Rua Pamplona 145, 01405-900 Sao Paulo, Brazil}

\end{center}
\vspace{1cm}

\begin{abstract}
\noindent
We present new theoretical results on the spectrum of the quantum
field theory of the Double Sine Gordon model. This non--integrable
model displays different varieties of kink excitations and bound
states thereof. Their mass can be obtained by using a semiclassical
expression of the matrix elements of the local fields. In certain
regions of the coupling--constants space the semiclassical method
provides a picture which is complementary to the one of the Form
Factor Perturbation Theory, since the two techniques give information
about the mass of different types of excitations. In other regions the
two methods are comparable, since they describe the same kind of
particles. Furthermore, the semiclassical picture is particularly
suited to describe the phenomenon of false vacuum decay, and it also
accounts in a natural way the presence of resonance states and the
occurrence of a phase transition.

\vspace{0.5cm}

\hrulefill

E-mail addresses: mussardo@sissa.it,\,\, riva@sissa.it,\,\, sotkov@ift.unesp.br

\end{abstract}

\end{titlepage}

\newpage

\section{Introduction}\label{intro}
As a natural development of the studies on integrable quantum field
theories, there has been recently an increasing interest in studying
the properties of non--integrable quantum field theories in $(1+1)$
dimensions, both for theoretical reasons and their application
to several condensed--matter or statistical systems. However,
contrary to the integrable models, many features of these quantum
field theories are still poorly understood: in most of the cases,
in fact, their analysis is only qualitative and even some of their
basic data, such as the mass spectrum, are often not easily
available. Although one could always rely on numerical methods to
shed some light on their properties, it would be obviously useful
to develop some theoretical tools to control them analitically. In
this respect, there has been recently some progress, thanks to two
different approaches.

The first approach, called the Form Factor Perturbation Theory
(FFPT) \cite{DMS,dm}, is best suited to deal with those
non--integrable theories close to the integrable ones. It permits,
in particular, to obtain quantitative predictions on their mass
spectrum, scattering amplitudes and other physical quantities. As
any other perturbation scheme, it works finely as far as the
non--integrable theory is an adiabatic deformation of the original
integrable model, i.e. when the two theories are isospectral. This
happens when the field which breaks the integrability is {\em
local} with respect to the operator which creates the particles.
If, on the contrary, the field which moves the theory away from
integrability is {\em non--local} with respect to the particles,
the resulting non--integrable model generally displays confinement
phenomena and, in this case, some caution has to be taken in
interpreting these perturbative results.

The second approach, known as Semiclassical Method and based on the seminal
work of Dashen, Hasslacher and  Neveu \cite{DHN}, is on the other hand best
suited to deal with those quantum field theories (integrable or not) having
kink excitations of large mass in their semiclassical limit. Under
these circumstances, in fact, once the non--perturbative classical
solutions are known, it is relatively simple to determine the
two--particle Form Factors on the kink states of the basic fields
of the theory and to extract the spectrum of the excitations from
their pole structure \cite{goldstone,Fateev,MRS}. Although this method
is restricted to work in a semiclassical regime, it permits however
to analyze non--integrable theories in the whole coupling--constants
space, even far from the integrable points.

An interesting non--integrable model where both approaches can
be used is the so--called Double Sine--Gordon Model (DSG). Its
Lagrangian density is given by
\EQ
{\cal L} \,=\,\frac{1}{2} (\partial_{\mu} \varphi)^2 -
V(\varphi) \,\,\,,
\label{lagrangian}
\EN
with
\begin{equation}\label{pot1}
V(\varphi) \,= \, -\frac{\mu}{\beta^{2}}\,\cos\beta\,\varphi -
\frac{\lambda}{\beta^{2}} \,\cos\left(\frac{\beta}{2}\,
\varphi+\delta\right) + C\;,
\end{equation}
where $C$ is a constant that has be chosen such that to have a
vanishing potential energy of the vacuum state. The classical
dynamics of this model has been extensively studied in the past
by means of both analytical and numerical techniques (see
\cite{sodano} for a complete list of the results), while its
thermodynamics has been studied in \cite{condat} by using the
transfer integral method \cite{transferintegral}.

With $\lambda$ or $\mu$ equal to zero, the DSG reduces to the
ordinary integrable Sine--Gordon (SG) model with frequency $\beta$
or $\beta/2$ respectively. Hence the DSG model with a small value
of one of the couplings can be regarded as a deformation of the
corresponding SG model and studied, therefore, by means of the
FFPT \cite{dm}. On the other hand, for $\beta \rightarrow 0$,
irrespectively of the value of the coupling constants $\lambda$
and $\mu$, the DSG model reduces to its semiclassical limit.
Despite the non--integrable nature of the DSG model, its classical
kink solutions are -- remarkably enough -- explicitly known
\cite{sodano,condat} and therefore the Semiclassical Method
can be successfully applied to recover the (semi--classical)
spectrum of the theory. As we will see in the following, the
two approaches turn out to be complementary in certain regions
of the coupling constants, i.e. both are needed in order to get
the whole mass spectrum of the theory, whereas in other regions
they provide the same picture about the spectrum of the excitations.

Apart from the theoretical interest in testing the efficiency of
the two methods on this specific model where both are applicable,
the study of the DSG is particularly important since this model plays
a relevant role in several physical contexts, either as a classical
non--linear system or as a quantum field theory. At the classical
level, its non--linear equation of motion can be used in fact to study
ultra--short optical pulses in resonant degenerate medium or
texture dynamics in He$^3$ (see, for instance, \cite{bullough}
and references therein). As a quantum field theory, depending on the
values of the parameters $\lambda,\mu, \beta,\delta$ in its Lagrangian,
it displays a variety of physical effects, such as the decay of
a false vacuum or the occurrence of a phase transition, the confinement
of the kinks or the presence of resonances due to unstable bound
states of excited kink--antikink states. Moreover, it finds interesting
applications in the study of several systems, such as the massive
Schwinger field theory or the Ashkin--Teller model \cite{dm}, as
well as in the analysis of the O(3) non--linear sigma model with
$\theta$ term \cite{cm}, i.e. the quantum field theory relevant
for understanding the dynamics of quantum spin chains
\cite{haldane,afflecklecture}. The DSG model also matters in
the investigation of other interesting condensed matter phenomena,
such as the soliton confinement of spin--Peierls antiferromagnets
\cite{affleck}, the dynamics of the spin chains in a staggered external
field or the electron interaction in a staggered potential \cite{fabr}.

Motivated by the above combined theoretical and physical interests,
a thorough study of the spectrum of the DSG model seems therefore
to be particularly interesting and in this paper we present the
results of such analysis.

The paper is organised as follows: in Section \ref{secFFPT} we
briefly recall the basic formulas of the Form Factor Perturbation
Theory whereas in Section \ref{secSM} we remind the basic results
of the Semiclassical Method. Section \ref{secDSG} is devoted to
the semiclassical analysis of the spectrum of the DSG model and
its comparison with the results coming from FFPT. Section
\ref{secfalse} deals with the analysis of false vacuum decay. In
Section \ref{secres} we discuss the occurrence of resonance
phenomena in the DSG in relation with analogous effects observed
in the classical scattering of kink states. Our conclusions are in
Section \ref{secconcl}. The paper also contains several
appendices. In Appendix \ref{kmasscorr} we compute the kink mass
corrections by using the FFPT, in Appendix \ref{secFF} we collect
the relevant expressions of the semiclassical Form Factors,
Appendix \ref{breathers} is devoted to the analysis of
neutral states in comparison with the Sine--Gordon model, and in
Appendix \ref{secDShG} we discuss the basic results in
a closely related model, i.e. the Double Sinh--Gordon model.

\section{Form Factor Perturbation Theory}\label{secFFPT}
\setcounter{equation}{0}
The method of the Form Factor Perturbation Theory (FFPT) \cite{DMS,dm}
permits to analyse a non--integrable quantum field theory when its
action ${\cal A}$ is represented by a deformation of an integrable one
${\cal A}_{0}$ through a given operator $\Psi$:
\begin{equation}
{\cal A}\, = \, {\cal A}_{0} + g\int d^{2}x\,\Psi(x)\;.
\end{equation}
One of the first consequences of moving away from integrability is
a change in the spectrum of the theory: the first order corrections
to the mass of the particle $a$ belonging to the spectrum of the
unperturbed theory is in fact given by
\begin{equation}
\label{FFPTmass}
\delta m_{a}^{2} \,= \, 2 g F_{a\bar{a}}^{\Psi}(i\pi)
+ O(g^{2})\;,
\end{equation}
where the particle--antiparticle Form Factor of the operator
$\Psi(x)$, defined by the matrix element\footnote{We adopt the
standard parameterization of the on--shell two--dimensional
momenta given in terms of the rapidity, $p^{(0)} = m \cosh\theta$,
$p^{(1)} = m \sinh\theta$.} \EQ F_{a \bar{a}}^{\Psi}(\theta_1 -
\theta_2) \,=\, \langle 0 \mid \Psi(0) \mid a(\theta_1)
\bar{a}(\theta_2) \rangle \,\,\,, \EN is introduced. The mass
correction (\ref{FFPTmass}) may be finite or divergent, depending
on the locality properties of the operator $\Psi(x)$ with respect
to the particle $a$. The situation was clarified in \cite{dm} and
it is worth recalling the main conclusion of that analysis.

In integrable theories, the Form Factors of a generic scalar
operator ${\cal O}(x)$ can be determined due to the simple form
assumed by the Watson equation \cite{KW,Smirnov} and for the
two--particle case, one has \EQ F_{a\bar{a}}^{\cal O}(\th) \,=\,
S_{a\bar{a}}^{b\bar{b}}(\th) F_{\bar{b}b}^{\cal O}(-\th)\,\,,
\label{uni} \EN \EQ F_{a\bar{a}}^{\cal O}(\th+2i\pi) \,=
\,e^{-2i\pi\gamma_{{\cal O},a}} F_{\bar{a}a}^{\cal O}(-\th)\,\,,
\label{cross} \EN where $\theta = \theta_1 - \theta_2$. In the
first equation, expressing the discontinuity of the matrix element
across the unitarity cut, $S_{a\bar{a}}^{b\bar{b}}(\th)$ is the
elastic two--body scattering amplitude. In the second equation,
expressing the crossing symmetry of the Form Factor, the explicit
phase factor $e^{-2i\pi \gamma_{{\cal O},a}}$ is inserted to take
into account a possible semi-locality of the operator which
interpolates the particle $a$ (i.e. any operator $\varphi_a$ such
that $\langle 0|\varphi_a|a\rangle\neq 0$) with respect to the
operator ${\cal O}(x)$ \footnote{Consistency of eq.\,(\ref{cross})
requires $\gamma_{{\cal O},\bar{a}} = - \gamma_{{\cal O},a}$.}.
When $\gamma_{{\cal O},a}  =0$, there is no crossing symmetric
counterpart to the unitarity cut but when $\gamma_{{\cal O},a}\neq
0$, there is instead a non-locality discontinuity in the plane of
the Mandelstam variable $s$, with $s=0$ as branch point\footnote{
The Mandelstam variable $s$ is expressed by $s = (p_a +
p_{\bar{a}})^2 = 4 m_a^2 \cosh^2(\theta/2)$.}. In the rapidity
parameterization there is however no cut because the different
Riemann sheets of the $s$-plane are mapped onto different sections
of the $\th$-plane; the branch point $s=0$ is mapped onto the
points $\th=\pm i\pi$ which become therefore the locations of
simple {\em annihilation} poles. The residues at these poles are
given by \cite{Smirnov} (see also \cite{Yurov}) \EQ -i\,{\mbox
Res}_{\th=\pm i\pi}F_{a\bar{a}}^{\cal O}(\th)= (1-e^{\mp
2i\pi\gamma_{{\cal O},a}})\langle 0|{\cal O}|0\rangle\,\,.
\label{pole} \EN In a Sine--Gordon model with frequency $\beta$,
an exponential operator $\Psi_{\alpha} = e^{i \alpha \varphi}$ has
a semi--locality index with respect to the soliton $s$ of the
theory given by $\gamma_{\alpha,s} = \alpha/\beta$ whereas it has
a vanishing semi--locality index with respect to the breather
particles \cite{dm}. This implies that, taking the Sine--Gordon
action as the integrable $A_0$ and $\Psi_{\alpha}$ as the
perturbing operator, the formula (\ref{FFPTmass}) can be safely
applied to compute the first order correction to the mass of the
breathers, whereas a divergence may appear in an analogous
computation of the mass correction of the solitons. This
divergence has to be seen as the mathematical signal that the
solitons of the original integrable model no longer survive as
asymptotic particles of the perturbed theory, i.e. they are
confined.

\section{Semiclassical Method}\label{secSM}
\setcounter{equation}{0}
The semiclassical quantization of a field theory defined by a
potential $V(\phi)$ consists in identifying a classical background
$\phi_{cl}(x)$, which satisfies the equation of motion
\begin{equation} \label{eom}
\partial_{\mu}\partial^{\mu}\phi_{cl} + V'(\phi_{cl}) \,=\, 0\;,
\end{equation}
and in applying to it various well established techniques, like
the path integral formalism \cite{pathint} or the solution of the
field equations in classical background \cite{DHN}, usually called
the DHN method (for a systematic review, see \cite{raj}).

The procedure is particularly simple and interesting if one
considers classical field solutions $\phi_{cl}(x)$ in $(1+1)$
dimensions which are static "kink" configurations interpolating
between degenerate minima of the potential, and whose quantization
gives rise to a particle-like spectrum.

A remarkable result, due to Goldstone and Jackiw \cite{goldstone}
(see also \cite{Fateev} for a non--relativistic context), is that the
classical background $\phi_{cl}(x)$ has the quantum meaning of Fourier
transform of the Form Factor of the basic field $\phi(x)$ between kink
states. The technique to derive this result relies on the Heisenberg
equation of motion satisfied by the quantum field $\phi(x)$ together
with the basic hypothesis that the kink momentum is very small compared
to its mass\footnote{The mass of kink state is inversely proportional to the
coupling constant, considered small in the semiclassical regime, and
therefore the kink is a heavy particle in this limit.}. In \cite{MRS},
we have refined the original argument overcoming its serious drawback
of being formulated non-covariantly in terms of the kink space-momenta.
This was possible thanks to the use of the rapidity variable $\theta$
of the kink states (and considering it as very small), instead of the
momentum.

The final result is the expression of the semiclassical form
factor between kink states as the Fourier transform of the
classical kink background, with respect to the Lorentz invariant
rapidity difference $\theta\equiv\theta_{1}-\theta_{2}$:
\begin{equation}
\label{ffinf}
<p_{1}|\phi(0)|p_{2}> \,\equiv\, f(\theta)\,\equiv \,
M_{cl}\, \int da\,e^{i\,M_{cl}\theta a} \phi_{cl}(a)\;,
\end{equation}
where $M_{cl}$ is the classical energy of the
kink\footnote{Along the same lines, it is possible to
prove that the form factor of an operator expressible as a
function of $\phi$ is given by the Fourier transform of the same
function of $\phi_{cl}$. For instance, the form factor of the
energy density operator $\varepsilon$ can be computed performing
the Fourier transform of $\varepsilon_{cl}(x) =
\frac{1}{2}\left(\frac{d\phi_{cl}}{dx}\right)^{2} + V[\phi_{cl}]$.}.
Having a covariant formulation, it is possible to express the
crossed channel form factor through the variable transformation
$\theta\rightarrow i\pi-\theta$:
\begin{equation}
\label{f2}
F_{2}(\theta) \,\equiv\, <0|\,\phi(0)|\,p_{1},\bar{p}_{2}> \, = \,
f(i\pi-\theta)\;.
\end{equation}
The analysis of this quantity provides a direct information about the
spectrum of the theory. Its dynamical poles, in fact, located at
$\theta^{*}=i(\pi-u)$ with $0<u<\pi$, coincide with the poles of
the kink--antikink $S$-matrix, and the relative bound states
masses can be then expressed as
\begin{equation}\label{bstmasses}
m_{(b)} \,= \, 2 M_{cl} \sin\frac{u}{2}\;.
\end{equation}
It is worth stressing that this procedure for extracting the
semiclassical bound states masses is remarkably simpler than the
standard DHN method of quantizing the corresponding classical
backgrounds, because in general these solutions depend also on
time and have a much more complicated structure than the kink
ones. Moreover, in non--integrable theories these backgrounds
could even not exist as exact solutions of the field equations:
this happens for example in the $\phi^{4}$ theory, where the DHN
quantization has been performed on some approximate backgrounds
\cite{DHN}.

\vspace{0.5cm}

In order to compute the first quantum corrections to the masses,
one has to quantize semiclassically the theory around the
classical solution by splitting the field as $\phi(x,t) =
\phi_{cl}(x) + \eta(x)e^{-i \omega t}$ and finding the eigenvalues
$\omega_{i}$ of the stability equation \cite{DHN,raj}
\begin{equation}\label{stability}
\left(-\partial_{x}^{2} + V''[\phi_{cl}(x)]\right) \,\eta_{i}(x) \,
= \, \omega_{i}^{2} \,\eta_{i}(x)\;.
\end{equation}
With these, the semiclassical energy levels are build as
\begin{equation}
{\cal E}_{\{n_{i}\}} \, = \, {\cal E}_{cl}
+ \hbar\sum_{i}\left(n_{i} + \frac{1}{2}\right)\omega_{i}+O(\hbar^{2})\;,
\end{equation}
and, in particular, the particles masses are given by the ground
state of these levels
\begin{equation}\label{e0}
{\cal E}_{0}\equiv{\cal E}_{\{n_{i}=0\}}\,=\,{\cal E}_{cl}
+ \frac{\hbar}{2}\sum_{i}\omega_{i}+O(\hbar^{2})\;.
\end{equation}
In the following, we will not include these corrections in our
results, since the analytical solution of the stability equation
(\ref{stability}) in the case of the DSG model is still missing.
Nevertheless, these corrections are not necessary in the cases in
exam, because we consider kink particles, for which the classical
energy is the term of leading order in the coupling, and their
bound states, for which expression (\ref{bstmasses}) already
encodes the first semiclassical corrections (see \cite{DHN}).

However, the eigenvalues $\omega_{i}$ play an important role in
the case of unstable particles, since the fingerprint of
instability is precisely the imaginary nature of some of these
frequencies. Hence, although we will obtain real values for the
masses of all the considered particles, we will always keep in
mind that many of these masses receive imaginary contributions
coming from some of the $\omega_{i}$.

\section{Semiclassical analysis of DSG particle spectrum}\label{secDSG}
\setcounter{equation}{0}
The double Sine-Gordon model is defined by the potential
\begin{equation}\label{pot}
V_{\delta}(\varphi)
\,=\, -\frac{\mu}{\beta^{2}}\,\cos\beta\,\varphi
- \frac{\lambda}{\beta^{2}}\,\cos\left(\frac{\beta}{2}\,
\varphi + \delta\right) + C \;,
\end{equation}
with the constant $C$ chosen such that the vacuum state has a vanishing
potential energy. We will study this theory in a regime of small $\beta$,
where the semiclassical results are expected to give a valuable approximation
of the spectrum\footnote{By applying the stability conditions found in
\cite{dm} to this model, they reduce to the condition $\beta^{2} <
8\pi$. Hence, for these values of $\beta$ and, in particular in
the semiclassical limit $\beta \to 0$, the potential (\ref{pot})
is stable under renormalization and no countertems have to be
added.}. At the quantum level, the different Renormalization Group
trajectories originating from the gaussian fixed point described
by the kinetic term $\frac{1}{2} (\partial_{\mu} \varphi)^2$ of
the lagrangian (\ref{lagrangian}) are labelled by the dimensionless
scaling variable $\eta = \lambda \mu^{-(8 \pi - \beta^2/4) /
(8\pi - \beta^2)}$ which simply reduces to the ratio $\eta =
\frac{\lambda}{\mu}$ in the semiclassical limit. When $\lambda$
or $\mu$ are equal to zero, the DSG model coincides with an ordinary
Sine-Gordon model with coupling $\beta$ or $\beta/2$, and mass scale
$\sqrt{\mu}$ or $\sqrt{\lambda/4}$, respectively.

Since for general values of the couplings the potential (\ref{pot})
presents a $\frac{4\pi}{\beta}$-periodicity, it was noticed in \cite{tak}
that one has an adiabatic perturbation of an integrable model only
if the $\lambda=0$ theory is regarded as a two--folded Sine-Gordon model.
This theory is a modification of the standard Sine-Gordon model, where
the period of the field $\phi$ is defined to be $\frac{4\pi}{\beta}$,
instead of $\frac{2\pi}{\beta}$ \cite{foldedSG}. As a consequence of this
new periodicity assignment, such a theory has two different degenerate
vacua $|k\,\rangle$, with $k=0,1$ and $|k+2\,\rangle\equiv|k\,\rangle$,
which are defined by $\langle\,k|\,\phi\,|\,k\,\rangle=\frac{2\pi}{\beta}
\,k$. Hence it has two different kinks, related to the classical backgrounds
by the formula
\begin{equation}\label{twofoldedkinks}
K_{k,k+1}^{cl}(x)\,=\,\frac{2k\pi}{\beta} +
\frac{4}{\beta}\arctan\,e^{\,m\,x}\,\qquad
k=0,1\;,
\end{equation}
and two corresponding antikinks, related to the classical solutions
by the expression
\begin{eqnarray}
K_{k+1,k}^{cl}(x)&=&\frac{2k\pi}{\beta} +
\frac{4}{\beta}\arctan\,e^{\,-m\,x}\\
&=&\frac{2(k+1)\pi}{\beta} -
\frac{4}{\beta}\arctan\,e^{\,m\,x}\,\qquad k=0,1\;.
\end{eqnarray}
Finally, in the spectrum there are also two sets of kink-antikink
bound states $b_{n}^{(l)}$, with $l=0,1$ and $n=1,...,
\left[\frac{8\pi}{\xi}\right]$.

The flow between the two limiting Sine-Gordon models (with frequency
$\beta$ or $\beta/2$, respectively) displays a variety of different
qualitative features, including confinement and phase transition
phenomena, depending on the signs of $\lambda$ and $\mu$, and on
the value of the relative phase $\delta$. However, it was observed
in \cite{dm} that the only values of $\delta$ which lead to inequivalent
theories are those given by $|\delta|\leq\frac{\pi}{2}$. Furthermore, in
virtue of the relations
\EQ
\begin{array}{rcl}
V_{\delta}\left(\phi
+ \frac{\pi}{\beta},\lambda,\mu\right) & = &
V_{\delta+\pi/2}\left(\phi,\lambda,-\mu\right)\;,\\
V_{\delta}(-\phi,\lambda,\mu) & = &
V_{-\delta}(\phi,\lambda,\mu)\;,
\end{array}
\EN
we can describe all the inequivalent possibilities keeping $\mu$
positive and the relative phase in the range $0 \leq \delta \leq
\frac{\pi}{2}$. The sign of the coupling $\lambda$, instead, simply
corresponds to a shift or a reflection of the potential, without
changing its qualitative features. As we are going to show in the
following, the case $\delta = \frac{\pi}{2}$ displays peculiar
features, while a common description is possible for any other
value of $\delta$ in the range $0 \leq \delta < \frac{\pi}{2}$.

\vspace{0.5cm}
In closing this discussion on the general properties of the DSG model,
we would like to mention that the possibility of writing exact classical
solutions for all the different kinds of topological objects in this
model finds a deep explanation in the relation between the trigonometric
potential (\ref{pot}) and power-like potentials. In fact, defining
\begin{equation}
\varphi\,=\,\frac{n\pi}{\beta}\pm \frac{4}{\beta}\arctan
Y\;,\qquad n=0,1,2,3\;,
\end{equation}
one can easily see that the first order equation which determines
the kink solution
\begin{equation}
\frac{1}{2}\left(\frac{d\varphi}{dx}\right)^{2} \,=\,
-\frac{\mu}{\beta^{2}}\,\cos\beta\,\varphi -
\frac{\lambda}{\beta^{2}}\,\cos\left(\frac{\beta}{2}\, \varphi +
\delta\right) + C
\end{equation}
is mapped into the equation for $Y$
\begin{equation}
\frac{1}{2}\left(\frac{dY}{dx}\right)^{2}\,=\,U(Y)\;,
\end{equation}
where $U(Y)$ describes various kinds of algebraic potentials,
depending on the values of $n$, $\delta$ and $C$. The $\delta=0$
case was analyzed in \cite{relationphi4} and its classical
solutions are very simple because $U(Y)$ only contains quartic
and quadratic powers of $Y$. It is easy to see that a similar
situation also occurs in the $\delta = \frac{\pi}{2}$ case; for
instance, choosing $n=1$ and $C=-\frac{1}{\beta^{2}}\left(\mu +
\frac{\lambda^{2}}{8\mu}\right)$, one obtains the quartic potential
\EQ
U(Y)\,=\, \frac{(4\mu + \lambda)^{2}}{128\mu}
\left(\frac{4\mu - \lambda}{4\mu + \lambda}\;Y^{2}-1\right)^{2}\;,
\EN
which has the well known classical background
\EQ
Y(x)\,=\,\sqrt{\frac{4\mu + \lambda}{4\mu - \lambda}}\,
\tanh\left(\sqrt{\mu-\frac{\lambda^{2}}{16\,\mu}}\;\frac{x}{2}\right)
\,\,\,.
\EN
For generic $\delta$, instead, also cubic and linear powers of $Y$
appear, making more complicated the analysis of the classical
solutions.

\subsection{$\delta=0$ case}

It is convenient to start our discussion with the case $\delta=0$.
This case, in fact, displays those topological features which are
common to all other models with $0 <\delta < \frac{\pi}{2}$, but
it admits a simpler technical analysis, due to the fact that parity
invariance survives the deformation of the original SG model. As we
will see explicitly, in this case the results of the FFPT and the
Semiclassical Method are complementary, since they describe
different kinds of excitations present in the theory.

\psfrag{phiopi}{$\frac{\phi}{\pi}$}

\begin{figure}[h]
\begin{tabular}{p{8cm}p{8cm}}
\psfrag{V}{$V(\mu=1,\,\lambda=1/2)$}
\psfig{figure=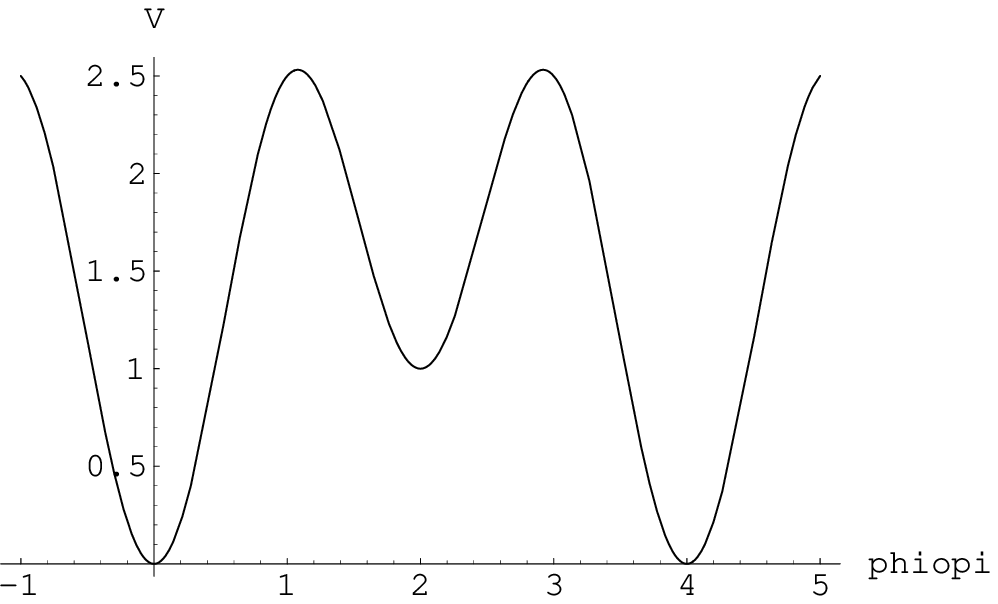,height=5cm,width=7cm}
& \psfrag{V}{$V(\mu=1,\,\lambda=4)$}
\psfig{figure=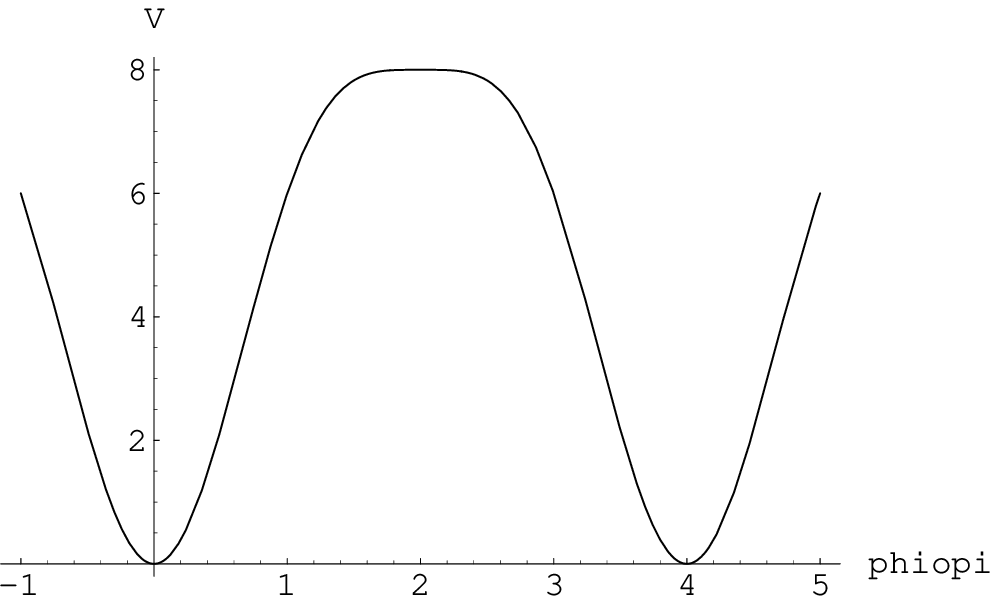,height=5cm,width=7cm}
\end{tabular}
\caption{DSG potential in the case $\delta=0$.}\label{figdelta0}
\end{figure}

Fig.\ref{figdelta0}  shows the shape of this DSG potential in the
two different regimes, i.e. (i) $0<\lambda<4 \mu$ and (ii) $\lambda>4\mu$.
The absolute minimum persists in the position $0\;(\text{mod}\,
\frac{4\pi}{\beta})$ for any values of the couplings, while the
other minimum at $\frac{2\pi}{\beta}\;(\text{mod}\,\frac{4\pi}{\beta})$
becomes relative and disappears at the point $\lambda=4\mu$. The breaking
of the degeneration between the two initial vacua in the
two--folded SG causes the confinement of the original SG solitons,
as it can be explicitly checked by applying the FFPT. The linearly
rising potential, responsible for the confinement of the SG solitons,
gives rise then to a discrete spectrum of bound states whose mass is
beyond $2 M_{SG}$, where $M_{SG}$ is the mass of the SG solitons
\cite{dm,affleck}.

The disappearing of the initial solitons represents, of course, a
drastic change in the topological features of the spectrum. At the
same time, however, a stable new static kink solution appears for
$\lambda\neq 0$, interpolating between the new vacua at $0$ and
$\frac{4\pi}{\beta}$. The existence of this new topological
solution is at the origin of the complementarity between the FFPT
and the Semiclassical Method. By the first technique, in fact, one
can follow adiabatically the deformation of the SG breathers
masses: these are neutral objects that persist in the theory
although the confinement of the original kinks has taken place. It
is of course impossible to see these particle states by using the
Semiclassical Method, since the corresponding solitons, which
originate these breathers as their bound states, have disappeared.
Semiclassical Method can instead estimate the masses of other
neutral particles, i.e. those which appear as bound states of the
new stable kink present in the deformed theory.

\begin{figure}[h]
\psfrag{x}{$x$}\psfrag{2 pi}{$2\pi$}\psfrag{4
pi}{$4\pi$}\psfrag{phiK(x)}{$\phi_{K}(x)$} \psfrag{- R}{$-R$}
\psfrag{R}{$R$}
\hspace{4cm}\psfig{figure=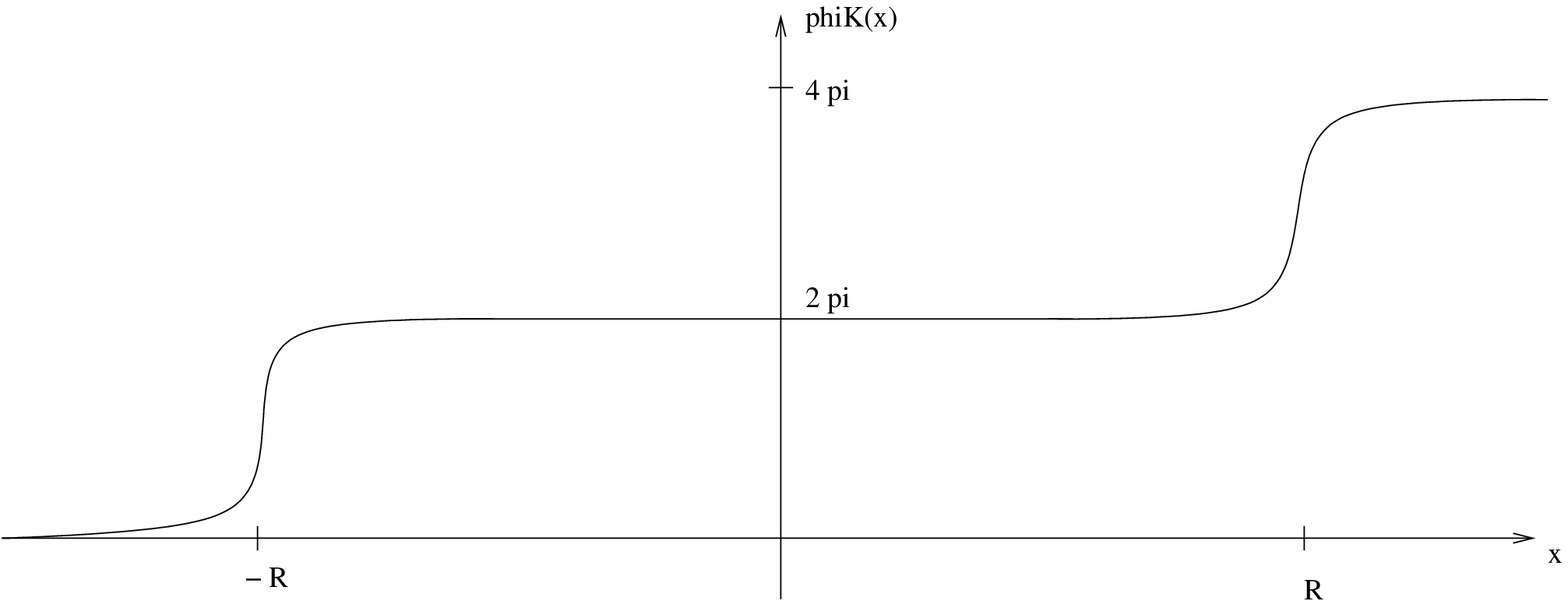,height=5cm,width=7cm}
\caption{Kink solution (\ref{doublekink})} \label{figdoublekink}
\end{figure}

This new kink solution, interpolating between $0$ and
$\frac{4\pi}{\beta}$, is given explicitly by
\begin{equation}\label{doublekink}
\varphi_{K}(x) = \frac{2\pi}{\beta}+
\frac{4}{\beta}\arctan\left[\sqrt{
\frac{\lambda}{\lambda+4\mu}}\,\sinh\left(m\,x\right)\right]\;,
\end{equation}
where
\begin{equation}\label{curvaturedeltazero}
m^{2}=\mu+\frac{\lambda}{4}ù
\end{equation}
is the curvature of the absolute minimum. Interestingly enough
\cite{sodano}, this background admits an equivalent expression in
terms of the superposition of two solitons of the unperturbed
Sine-Gordon model, centered at the fixed points $\pm R$
\begin{equation}
\varphi_{K}(x)=\varphi_{\text{SG}}(x+R)+\varphi_{\text{SG}}(x-R)\;,
\end{equation}
where $\varphi_{\text{SG}}(x) =
\frac{4}{\beta}\arctan\left[e^{m\,x}\right]$ are the usual
Sine-Gordon solitons with the deformed mass parameter
(\ref{curvaturedeltazero}) whereas their distance $2R$ is
expressed in terms of the couplings by
\[
R \, =\, \frac{1}{m}\,\text{arccosh}\sqrt{\frac{4\mu}{\lambda}+1}\,\,\,.
\]
By looking at Fig.\ref{figdoublekink}, it is clear that this background,
in the small $\lambda$ limit, describes the two confined solitons of SG,
which become free in the $\lambda=0$ point, i.e. where $R\to\infty$.

The classical energy of this kink is given by
\begin{equation}\label{doublekinkmass}
M_{K} \,= \, \frac{16\,m}{\beta^{2}}\,
\left\{1+\frac{\lambda}{\sqrt{4\mu(\lambda+4\mu)}}\,
\text{arctanh} \sqrt{\frac{4\mu}{\lambda+4\mu}}\right\}\;,
\end{equation}
and in the $\lambda\to 0$ limit it tends to twice the classical
energy of the Sine-Gordon soliton, i.e.
\begin{equation}
M_{K}\;{\mathrel{\mathop{\kern0pt\longrightarrow}\limits_{\lambda\to
0 }}}\;\frac{16\sqrt{\mu}}{\beta^{2}}\;,
\end{equation}
therefore confirming the above picture. In the $\mu\to 0$ limit,
the asymptotic value of the above expression is instead the mass
of the soliton in the Sine-Gordon model with coupling $\beta/2$.
The expansion for small $\mu$
\begin{equation}
\label{doublekinkfirst}
M_{K}\;{\mathrel{\mathop{\kern0pt\longrightarrow}
\limits_{\mu\to 0 }}}\;
\frac{8\sqrt{\lambda/4}}{(\beta/2)^{2}}
+ \frac{\mu}{\beta^{2}}\,\frac{32}{3\sqrt{\lambda}} +
O(\mu^{2})\;,
\end{equation}
gives the first order correction which is in agreement with the result
of the FFPT in the semiclassical limit (see Appendix \ref{kmasscorr}).

The bound states created by the kink (\ref{doublekink}) and
its antikink can be obtained by looking at the poles of the
semiclassical Form Factors of the fields $\varphi(x)$ and
$\varepsilon(x)$, reported in Appendix \ref{secFF}, and their
mass are given by\footnote{Due to parity invariance, the dynamical
poles of the form factor of $\varphi$ between kink states only give
the bound states with $n$ odd. The even states can be obtained from the
form factor of the energy operator $\epsilon(x)$.}
\begin{equation}
\label{doubleboundstates}
m_{(K)}^{(n)} \, = \, 2M_{K}\sin\left(n\,\frac{m}{2M_{K}}\right)
\,\,\,\,\,\,\,
,
\,\,\,\,\,\,\,
0 < n < \pi\frac{M_{K}}{m}\;.
\end{equation}
For small $\mu$ we easily recognize the perturbation of the
standard breathers in Sine-Gordon with $\beta/2$:
\begin{equation}\label{breathersmu}
m_{(K)}^{(n)}\;{\mathrel{\mathop{\kern0pt\longrightarrow}
\limits_{\mu\to 0}}}\;
\frac{64}{\beta^{2}}\sqrt{\frac{\lambda}{4}}
\sin\left(n\,\frac{\beta^{2}}{64}\right)
+ \frac{2}{3}\,\frac{\mu}{\sqrt{\lambda}}
\left[\frac{32}{\beta^{2}}\sin\left(n\,
\frac{\beta^{2}}{64}\right)+n\,\cos\left(n\,
\frac{\beta^{2}}{64}\right) \right]+O(\mu^{2})\;,
\end{equation}
while the expansion of the bound states masses for small $\lambda$
\begin{eqnarray*}
m_{(K)}^{(n)}\;&{\mathrel{\mathop{\kern0pt\longrightarrow}
\limits_{\lambda\to 0 }}}\;&
\frac{32\sqrt{\mu}}{\beta^{2}}\sin\left(n\,
\frac{\beta^{2}}{32}\right)
+ \\
&& +\frac{1}{8}\,\frac{\lambda}{\sqrt{\mu}}
\left[\left(1-\ln\frac{\lambda}{16\mu}\right)
\frac{32}{\beta^{2}}\sin\left(n\,\frac{\beta^{2}}{32}\right) +
n\,\ln\frac{\lambda}{16\mu}\,\cos\left(n\,
\frac{\beta^{2}}{32}\right) \right]+O(\lambda^{2})\;
\end{eqnarray*}
deserves further comments: in fact, although the above masses have
well-defined asymptotic values, they do not correspond however
to any state of the unperturbed SG theory. The reason is that
the classical background (\ref{doublekink}) in the $\lambda\to 0$
limit does not describe any longer a localized single particle. This
implies that its Fourier transform cannot be interpreted as the
two-particle Form Factor and, consequently, its poles cannot be
associated to any bound states.

A technical signal of the disappearing of the above mentioned
bound states in the $\lambda\to 0$ limit can be found by computing
the three particle coupling among the kink, the antikink and the
lightest bound state. The residue of the kink-antikink form factor
on the pole corresponding to the lightest bound state $b^{(1)}$
has to be proportional to the one-particle form factor
$<0|\,\phi|\,b^{(1)}>$ through the semiclassical 3-particle
on-shell coupling of kink, antikink and elementary boson
$g_{k\bar{k}b}$:
\begin{equation}
\textrm{Res}\,_{\theta=\theta_{1}}F_{2}(\theta) =
i\,\frac{g_{k\bar{k}b}}{2\sqrt{2}M_{\infty}\, m_{b}^{(1)} }\,
<0|\,\phi|\,b^{(1)}> \,\,\,.
\end{equation}
Since the one-particle form factor takes the constant value
$1/\sqrt{2}$, at leading order in $\beta$ we get
\begin{equation}
g_{K\,\bar{K}\,b}=\frac{1}{4\sqrt{\lambda}}\left(\frac{16
m}{\beta}\right)^{3}\left\{1+\frac{\lambda} {\sqrt{4\mu
(\lambda+4\mu)}}\,\text{arctanh}\sqrt{\frac{4\mu}
{\lambda+4\mu}}\right\}\;.
\end{equation}
The divergence of the coupling as $\lambda\to 0$ indicates that
the considered scattering processes cannot be seen anymore as a
bound state creation, i.e. the corresponding bound state
disappears from the theory. A general discussion of the same
qualitative phenomenon for the ordinary Sine-Gordon model can be
found in \cite{imcoupl}, where the disappearing from the theory of
a heavy breather at specific values of $\beta$ is explicitly
related to the divergence or to the imaginary nature of the three
particle coupling among this breather and two lightest ones.

\vspace{0.5cm}

\begin{figure}[h]
\hspace{4cm}\psfig{figure=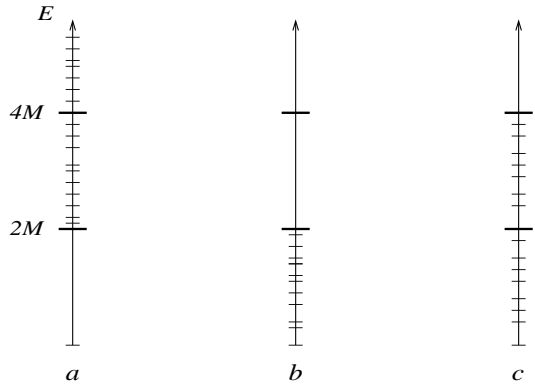,height=5cm,width=7cm}
\normalsize\caption{Neutral states coming from: a) solitons
confinement, b) deformations of SG breathers, c) bound states of
the kink (\ref{doublekink})} \label{figneutral}
\end{figure}

Summarizing, in this model we have three kinds of neutral objects,
i.e. meson particles. The first kind $(a)$ is given by the bound
states originating from the confinement potential of the original
solitons. These discrete states have masses above the threshold
$2M_{SG}$, where $M_{SG}$ is the mass of the SG solitons, and
merge in the continuum spectrum of the non-confined solitons in
the $\lambda\to 0$ limit \cite{dm,affleck}. The second kind $(b)$
is represented by the deformations of SG breathers, that can be
followed by means of the FFPT and have masses, for small
$\lambda$, in the range $[0,2M_{SG}]$. Finally, the third kind
$(c)$ is given by the bound states (\ref{doubleboundstates}) of
the stable kink of the DSG theory and they have masses in the
range $[0,4M_{SG}]$. All these mass spectra are drawn in Fig.
\ref{figneutral}.

Obviously, due to the non--integrable nature of this quantum field
theory not all these particles belong to the stable part of its
spectrum. Apart from a selection rule coming from the conservation
of parity, decay processes are expected to be simply controlled by
phase--space considerations, i.e. a heavier particle with mass
$M_h$ will decay in lighter particles of masses $m_i$ satisfying
the condition \EQ M_h \geq \sum_i m_i \,\,\,.
\label{stabilitythreshold} \EN Hence, to determine the stable
particles of the theory, one has initially to identify the
lightest mesons of odd and even parity with mass $m^{\star}_-$ and
$m^{\star}_+$ ($m^{\star}_- < m ^{\star}_+$), respectively. Then,
the stable particles of even parity are those with mass $m$ below
the threshold $2 m^{\star}_-$ whereas the stable particles of odd
parity are those with mass $m < m^{\star}_- + m^{\star}_+$. For
instance, in the $\mu\to 0$ limit we know that the only stable
mesons are those given by the particles $(c)$, as confirmed by the
expansion (\ref{breathersmu}). Hence, in this limit no one of the
other neutral particles is present as asymptotic states. For the
mesons of type $(a)$, this can be easily understood since they are
all above the threshold dictated by the lightest neutral particle.
The situation is more subtle, instead, for the states $(b)$.
However, their absence in the theory with $\mu \to 0$ clearly
indicates that at some particular value of $\lambda$ even the
lightest of these objects acquires a mass above the threshold
$2m_{(K)}^{(1)}$, with $m_{(K)}^{(1)}$ given by
(\ref{doubleboundstates}). Analogous analysis can be done for
other values of the couplings so that the general conclusion is
that most of the above neutral states are nothing else but
resonances of the DSG model.

\vspace{0.5cm}

In addition to the above scenario of kink states and bound state
thereof, in the region $\lambda < 4 \mu$ there is another
non-trivial static solution of the theory, defined over the false
vacuum placed at $\varphi = \frac{2\pi}{\beta}$. It interpolates
between the two values $\frac{2\pi}{\beta}$ and
$\frac{4\pi}{\beta} - \frac{2}{\beta}\arccos(1-\lambda/2\mu)$, and
then it comes back. Its explicit expression is given by
\begin{equation}
\label{bounce}
\varphi_{B}(x) \,=\,\frac{4\pi}{\beta}-
\frac{4}{\beta}\arctan\left[\sqrt{\frac{\lambda}
{4\mu-\lambda}}\,\cosh\left(m_{f}\,x\right)\right]\;,
\end{equation}
where
\begin{equation}
\label{falsecurvature}
m^{2}_{f} \,=\, \mu-\frac{\lambda}{4} \;
\end{equation}
is the curvature of the relative minimum. Similarly to the kink
(\ref{doublekink}), it admits an expression in terms of a soliton
and an antisoliton of the unperturbed SG model:
\begin{equation}
\label{bouncesumSG}
\varphi_{B}(x)\,=\, \varphi_{\text{SG}}(x+R)
+ \varphi_{\text{SG}}(-(x-R))\;,
\end{equation}
where now $\varphi_{\text{SG}}(x) = \frac{4}{\beta}\arctan
\left[e^{m_{f}\,x}\right]$ are the Sine-Gordon solitons with
the deformed mass parameter (\ref{falsecurvature}) whereas
their distance $2R$ is now given by
\begin{equation}
\label{fixedR}
R \,=\, \frac{1}{m_{f}}\,\text{arcsinh}\sqrt{\frac{4\mu}
{\lambda}-1}\,\,\,.
\end{equation}

\begin{figure}[h]
\psfrag{x}{$x$}\psfrag{2 pi}{$2\pi$}\psfrag{4
pi}{$4\pi$}\psfrag{phiB(x)}{$\phi_{B}(x)$} \psfrag{R}{$R$}
\psfrag{- R}{$-R$}
\hspace{4cm}\psfig{figure=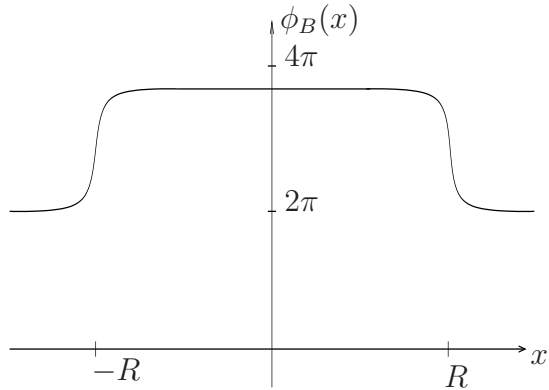,height=5cm,width=7cm}
\caption{Bounce-like solution (\ref{bounce})}\label{figbounce}
\end{figure}

In the small $\lambda$ limit, it is clear that this background describes
the confined soliton and antisoliton of the SG model, which become free
in the $\lambda=0$ point, i.e. where $R\to\infty$.

The classical background (\ref{bounce}) is not related to any stable
particle in the quantum theory. This can be directly seen from equation
(\ref{stability}); in fact, Lorentz invariance
always implies the presence of the eigenvalue $\omega_{0}^{2}=0$, with
corresponding eigenfunction $\eta_{0}(x) = \frac{d}{dx}\varphi_{cl}(x)$.
However, in the case of the solution (\ref{bounce}) the eigenfunction
$\eta_{0}$ clearly displays a node, which indicates that the corresponding
eigenvalue is not the smallest in the spectrum. Hence, there must be a
lower eigenvalue $\omega_{-1}^{2}<0$, with a corresponding imaginary
part of the mass relative to this particle state. Furthermore, the
instability of (\ref{bounce}) can be related to the theory of false
vacuum decay \cite{coleman,langer}: due to the deep physical interest
of this topic, we will discuss it separately in Section \ref{secfalse}.

\subsection{Comments on generic $\delta$ case}

\psfrag{phiopi}{$\frac{\phi}{\pi}$}

\begin{figure}[h]
\begin{tabular}{p{8cm}p{8cm}}
\psfrag{V}{$V(\mu=1,\,\lambda=1)$}
\psfig{figure=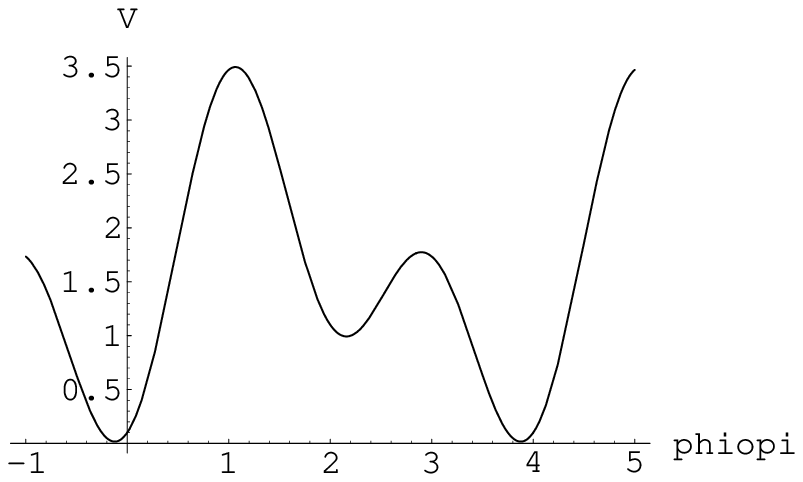,height=5cm,width=7cm} &
\psfrag{V}{$V(\mu=1,\,\lambda=2.3)$}
\psfig{figure=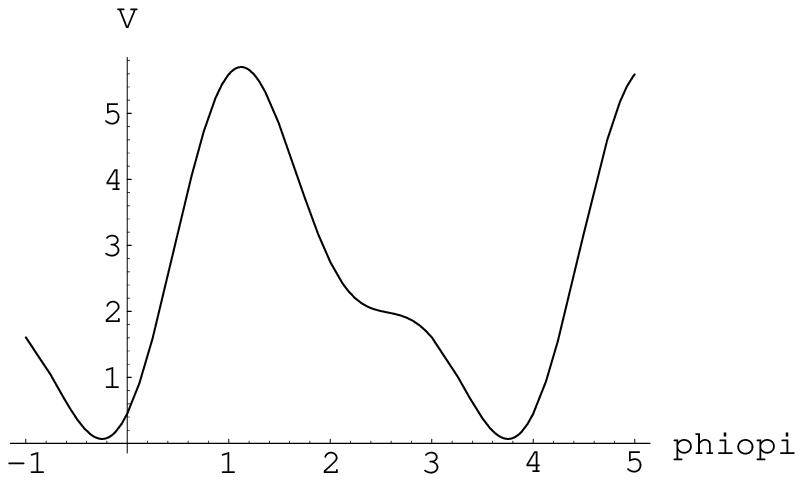,height=5cm,width=7cm}
\end{tabular}
\caption{DSG potential in the case
$\delta=\frac{\pi}{3}$.}\label{figdeltapi3}
\end{figure}

We have already anticipated that the qualitative features of the
theory relative to $\delta=0$ case are common to all other theories
associated to the values of $\delta$ in the range $0<\delta<\frac{\pi}{2}$.
This can be clearly understood by looking at the shape of the potential,
which is shown in Fig.\,\ref{figdeltapi3} for the case $\delta=\frac{\pi}{3}$.

In contrast to the $\delta=0$ case, parity invariance is now lost
in these models, and the minima move to values depending on the
couplings. Furthermore, in addition to the change in the nature
of the original vacuum at $\frac{2\pi}{\beta}$, which becomes a
relative minimum by switching on $\lambda$, there is also a lowering
of one of the two maxima. These features make much more complicated
the explicit derivation of the classical solutions, as we have mentioned
at the beginning of the Section.

However, it is clear from Fig. \ref{figdeltapi3} that the excitations
of these theories share the same nature of the ones in the $\delta=0$
case. In fact, the original SG solitons undergo a confinement, while a new
stable topological kink appears, interpolating between the new degenerate
minima. Hence, the analysis performed for $\delta = 0$ still holds in
its general aspects, i.e. also in these cases the spectrum consists of a
kink, antikink, and three different kinds of neutral particles.

\subsection{$\delta = \frac{\pi}{2}$ case}

The value $\delta = \frac{\pi}{2}$ describes the peculiar case in
which no confinement phenomenon takes place, since the two
different vacua of the original two--folded SG remain degenerate
also in the perturbed theory. As a consequence, the original SG
solitons are also asymptotic states in the perturbed theory.
By means of the Semiclassical Method we can then compute their
bound states, which represent the deformations of the two sets
of breathers in the original two--folded SG. Hence, in this specific
case FFPT and semiclassical method describe the same objects, and
their results can be compared in a regime where both $\beta$ and
$\lambda$ are small.

\footnotesize

\psfrag{phiopi}{$\frac{\phi}{\pi}$}

\begin{figure}[h]
\begin{tabular}{p{8cm}p{8cm}}
\psfrag{V}{$V(\mu=1,\,\lambda=1)$}\psfig{figure=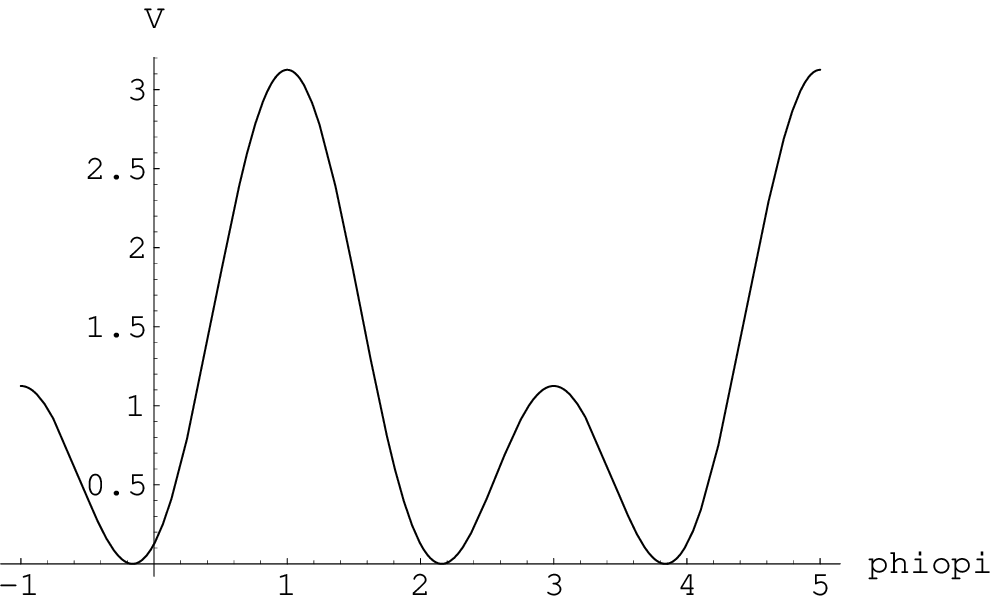,
height=5cm,width=7cm} \vspace{0.2cm}&
\psfrag{V}{$V(\mu=1,\,\lambda=4)$}
\psfig{figure=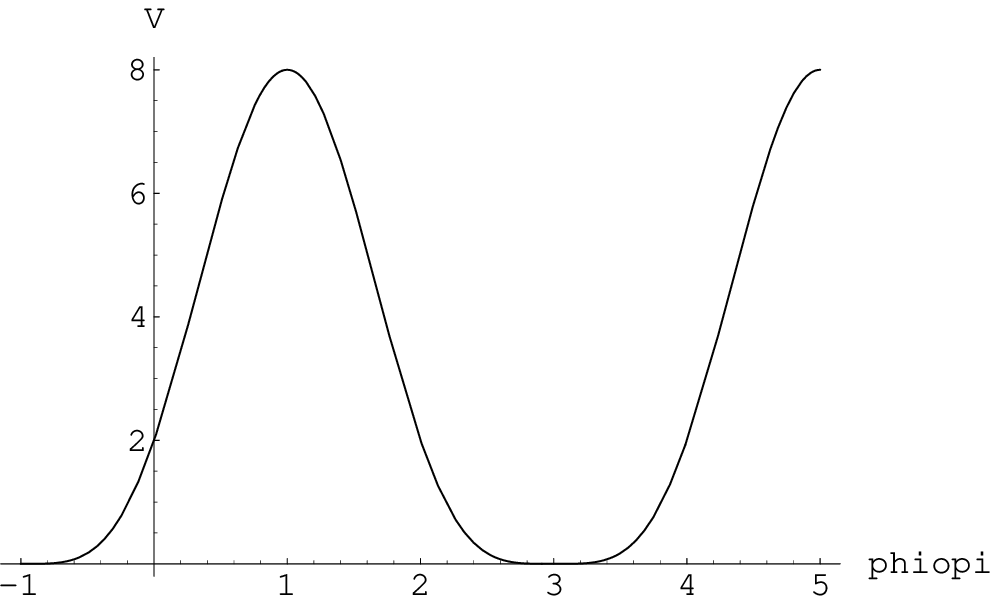,height=5cm,width=7cm}
\end{tabular}
\caption{DSG potential in the $\delta=\frac{\pi}{2}$
case.}\label{figdeltapi2}
\end{figure}

\normalsize

Fig.\,\ref{figdeltapi2} shows the behavior of this DSG potential.
There are two regions, qualitatively different, in the space of
parameters, the first given by $0 < \lambda < 4 \mu$ and the
second given by $\lambda > 4 \mu$. They are separated by the value
$\lambda = 4 \mu$ which has been identified in \cite{dm} as a phase
transition point. We will explain how this identification is confirmed
in our formalism.

Let's start our analysis from the coupling constant region where
$\lambda < 4 \mu$. Switching on $\lambda$, the original
inequivalent minima of the two--folded Sine-Gordon, located at
$\phi_{\text{min}} = 0,\,\frac{2\pi}{\beta}\;
(\text{mod}\,\frac{4\pi}{\beta})$, remain degenerate and move to
$\phi_{\text{min}} = -\phi_{0},\,\frac{2\pi}{\beta} +
\phi_{0}\;(\text{mod}\,\frac{4\pi}{\beta})$, with
$\phi_{0}=\frac{2}{\beta}\arcsin\frac{\lambda}{4\mu}$. The common
curvature of these minima is
\begin{equation}\label{curvaturedeltapi2}
m^{2}\,=\, \mu-\frac{1}{16}\,\frac{\lambda^{2}}{\mu} \;.
\end{equation}
Correspondingly there are two different types of kinks, one called
\lq\lq large kink" and interpolating through the higher barrier
between $-\phi_{0}$ and $\frac{2\pi}{\beta} + \phi_{0}$, the other
called \lq\lq small kink" and interpolating through the lower
barrier between $\frac{2\pi}{\beta} + \phi_{0}$ and
$\frac{4\pi}{\beta} - \phi_{0}$. Their classical expressions are
explicitly given by
\begin{equation}\label{largekink}
\varphi_{L}(x)\,=\, \frac{\pi}{\beta}+
\frac{4}{\beta}\arctan\left[\sqrt{\frac{
4\mu+\lambda}{4\mu-\lambda}}\,\tanh\left(
\frac{m}{2}\,x\right)\right]\qquad(\text{mod}\; 4\pi)\;,
\end{equation}
\begin{equation}\label{smallkink}
\varphi_{S}(x)\,=\,\frac{3\pi}{\beta} +
\frac{4}{\beta}\arctan\left[\sqrt{\frac{ 4\mu - \lambda}{4\mu +
\lambda}}\,\tanh\left(\frac{m}{2}
\,x\right)\right]\qquad(\text{mod}\; 4\pi)\;.
\end{equation}
With the notation previously introduced, the vacuum structure of
the corresponding quantum field theory consists of two sets of
inequivalent minima, denoted by $\mid 0 \,\rangle $ and $\mid 1
\,\rangle$, identified modulo $2$, i.e. $\mid a + 2 n \, \rangle
\equiv \mid a \rangle$. The spontaneous breaking of the symmetry
$T:\,\varphi\,\to\, 2\pi-\varphi$ selects one of these minima as
the vacuum. If we choose to quantize the theory around $\mid 0
\,\rangle $, the admitted quantum kink states are $\mid L
\,\rangle = \mid K_{0,1}\,\rangle$ and $\mid \overline S \,\rangle
= \mid K_{0,-1}\, \rangle $, with the corresponding antikink
states $\mid \overline L \,\rangle =\mid K_{1,0} \,\rangle$ and
$\mid S \,\rangle = \mid K_{-1,0}\,\rangle$, and topological
charges \EQ
\begin{array}{c}
Q_{L} = - Q_{\overline L} = 1+\frac{\beta\phi_{0}}{\pi}\;,\\
Q_{S} = -Q_{\overline S} = 1-\frac{\beta\phi_{0}}{\pi}\;. \\
\end{array}
\label{topologicalcharge} \EN Multi--kink states of this theory
satisfy the selection rule coming from the continuity of vacuum
indices and are generically given by \EQ \mid K_{\alpha_1
\alpha_2}(\theta_1) \, K_{\alpha_2 \alpha_3}(\theta_2) \, \cdots
K_{\alpha_{n-2} \alpha_{n-1}}(\theta_{n-2}) \, K_{\alpha_{n-1}
\alpha_{n}}(\theta_{n-1}) \,\rangle \EN The leading contributions
to the masses of the large and small kink are given by their
classical energies, which can be easily computed
\begin{equation}\label{largesmallkinkmass}
M_{L,S}\,=\,\frac{8\,m}{\beta^{2}}
\left\{1\pm\frac{\lambda}{\sqrt{16\,\mu^{2}-\lambda^{2}}}
\left(\frac{\pi}{2}\pm\arcsin\frac{\lambda}{4\mu}\right)\right\} \,\,\,.
\end{equation}
The expansion of this formula for small $\lambda$ is given by
\begin{equation}\label{largesmallkinkfirst}
M_{L,S}\;{\mathrel{\mathop{\kern0pt\longrightarrow}
\limits_{\lambda\to 0 }}}\;
\frac{8\sqrt{\mu}}{\beta^{2}}\pm\frac{\lambda}{\beta^{2}}\,
\frac{\pi}{\sqrt{\mu}}+O(\lambda^{2})\;,
\end{equation}
and the first order correction in $\lambda$ coincides with the
result of FFPT in the semiclassical limit (see Appendix \ref{kmasscorr}).

Since two different types of kink $|\,L\rangle$ and $|\,S\rangle$ are
present in this theory, one must be careful in applying eq.\,(\ref{ffinf})
to recover the form factors of each kink separately. In fact, one could expect
that both types of kink contribute to the expansion over intermediate states used
in \cite{goldstone} to derive the result. For instance, starting from the vacuum
$\mid 0\,\rangle$ located at $\phi_{\text{min}} = -\phi_{0}$ there might be
the intermediate matrix elements $_0\langle \bar{S} \mid {\cal O} \mid L
\rangle_0$ and $_0\langle L \mid {\cal O} \mid \bar{S} \rangle_0$. However,
if ${\cal O}$ is a non--charged local operator, it easy to see that these
off-diagonal elements have to vanish for the different topological charges of
$|\,L\rangle$ and $|\,S\rangle$. Hence, the expansion over intermediate states
diagonalizes and one recovers again eq.\,(\ref{ffinf}).

Therefore, from the dynamical poles of the form factor of $\varphi$ on the large
and small kink-antikink states, reported in Appendix \ref{secFF}, we can extract
the semiclassical masses of two sets of bound states:
\begin{equation}
\label{largebound}
m_{(L)}^{(n)}\,=\, 2 M_{L} \sin\left(n_{L}\,\frac{m}{2M_{L}}\right)
\,\,\,\,\,\,\,
,
\,\,\,\,\,\,\,
0 < n_{L} < \pi\frac{M_{L}}{m}\;,
\end{equation}
\begin{equation}\label{smallbound}
m_{(S)}^{(n)}\, =\, 2 M_{S} \sin\left(n_{S}\,\frac{m}{2M_{S}}\right)
\,\,\,\,\,\,\,
,
\,\,\,\,\,\,\,
0 < n_{S} < \pi\frac{M_{S}}{m}\;.
\end{equation}
Expanding for small $\lambda$, we can see that these states
represent the perturbation of the two sets of breathers in the
original two--folded Sine-Gordon model:
\begin{equation}
\label{linearlambda}
m_{(L,S)}^{(n)}\;{\mathrel{\mathop{\kern0pt\longrightarrow}
\limits_{\lambda\to 0 }}}\;\frac{16\sqrt{\mu}}{\beta^{2}}
\sin\left(n\,\frac{\beta^{2}}{16}\right)\pm
2\pi\frac{\lambda}{\sqrt{\mu}}
\left[\frac{1}{\beta^{2}}\sin\left(n\,
\frac{\beta^{2}}{16}\right) - \frac{n}{16}\cos\left(n\,
\frac{\beta^{2}}{16}\right)
\right] + O(\lambda^{2})
\end{equation}
A discussion of these results, in comparison with previous studies
of this model \cite{tak}, is reported in Appendix C.

Concerning the stability of the above spectrum, for $\lambda <
4 \mu$ the only stable bound states are the ones with $m_{(L,S)}^{(n)}
< 2 m_{(S)}^{(1)}$; for $\lambda$ close enough to $4\mu$, however,
the small kink creates no bound states, hence the stability condition
becomes $m_{(L)}^{(n)} < 2 m_{(L)}^{(1)}$.

\vspace{0.5cm}

In the limit $\lambda\to 4\mu$, $\phi_{0}$ tends to
$\frac{\pi}{\beta}$, the two minima at $\frac{2\pi}{\beta} +
\phi_{0}$ and $\frac{4\pi}{\beta} - \phi_{0}$ coincide and the
small kink disappears, becoming a constant solution with zero
classical energy. All the large kink bound states masses collapse
to zero, and in this limit all dynamical poles of the large kink
form factor disappear. This is nothing else but the semiclassical
manifestation of the occurrence of the phase transition present in
the DSG model (see \cite{dm}).

\vspace{0.5cm}

In the second coupling constant region, parameterized by $\lambda > 4 \mu$,
there is only one minimum at fixed position $-\frac{\pi}{\beta}\;(\text{mod}\,
\frac{4\pi}{\beta})$, with curvature
\begin{equation}\label{curvaturedeltapi2sec}
m^{2}\,=\,\frac{\lambda}{4}-\mu\;.
\end{equation}
There is now only one type of kink, given by
\begin{equation}\label{commonkink}
\varphi_{K}(x)\,=\,\frac{\pi}{\beta}+
\frac{4}{\beta}\arctan\left[\sqrt{\frac{\lambda}{\lambda-4\mu}}
\,\sinh\left(m\,x\right)\right] \,\,\,.
\end{equation}
Its classical mass, expanded for small $\mu$, is again in
agreement with FFPT (see Appendix \ref{kmasscorr}):
\begin{eqnarray}\label{commonkinkmass}
&&M_{K}\,=\,\frac{16\,m}{\beta^{2}}\,\left\{1
+ \frac{\lambda}{4\sqrt{\mu(\lambda-4\mu)}}\left(\frac{\pi}{2} -
\arcsin\frac{\lambda - 8\mu}{\lambda}\right)\right\}\;\to\\
&&\quad\;{\mathrel{\mathop{\kern0pt\longrightarrow}
\limits_{\lambda\to 0 }}}\;\;
\frac{8\sqrt{\lambda/4}}{(\beta/2)^{2}}
- \frac{\mu}{\beta^{2}}\,\frac{32}{3\sqrt{\lambda}}+O(\mu^{2})\;.
\end{eqnarray}
The bound states of this kink (see Appendix \ref{secFF} for the
explicit Form Factor) have masses
\begin{equation}
m_{(K)}^{(n)}\,=\, 2M_{K}\sin\left(n\,\frac{m}{2M_{K}}\right)
\,\,\,\,\,\,\,
,
\,\,\,\,\,\,\,
0 < n < \pi\frac{M_{K}}{m}\;.
\end{equation}
For small $\mu$, these states are nothing else but the
perturbed breathers of the Sine-Gordon model with coupling
$\beta/2$:
$$
m_{(K)}^{(n)}\;{\mathrel{\mathop{\kern0pt\longrightarrow}
\limits_{\mu\to 0 }}}\;\frac{64}{\beta^{2}}\sqrt{\frac{\lambda}{4}}
\,\sin\left(n\,\frac{\beta^{2}}{64}\right)-\frac{2}{3}\,
\frac{\mu}{\sqrt{\lambda}}
\left[\frac{32}{\beta^{2}}\sin\left(n\,\frac{\beta^{2}}{64}\right)
+ n\,\cos\left(n\,
\frac{\beta^{2}}{64}\right) \right] + O(\mu^{2})
$$

\vspace{0.5cm}

In closing the discussion of the $\delta = \pi/2$ case, it is
interesting to mention another model which presents a similar
phase transition phenomenon, although in a reverse order. This
is the Double Sinh--Gordon Model (DShG), discussed in Appendix \ref{secDShG}.
The similarity is due to the fact that also in this case a
topological excitation of the theory becomes massless at the
phase transition point, but the phenomenon is reversed, because
in DSG the small kink disappears when $\lambda$ reaches the critical
value, while in DShG a topological excitation appears at some value
of the perturbing coupling.

\section{False vacuum decay}\label{secfalse}
\begin{figure}[h]
\footnotesize\psfrag{phi}{$\varphi$}\psfrag{U(phi)}
{$U(\varphi)$}\psfrag{phi+}{$\varphi_{+}$}
\psfrag{phi-}{$\varphi_{-}$} \psfrag{phi1}{$\varphi_{1}$}
\hspace{4cm}\psfig{figure=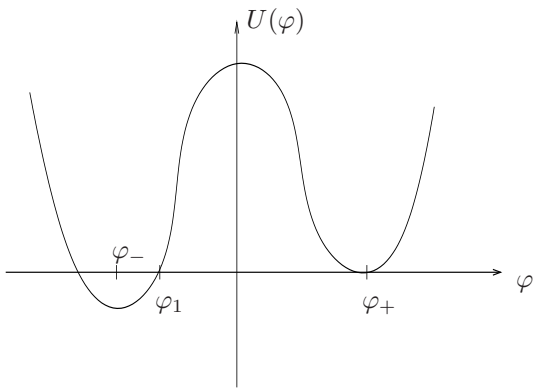,height=5cm,width=7cm}
\normalsize\caption{Generic potential for a theory with a false
vacuum}\label{figfalsevacpot}
\end{figure}

\setcounter{equation}{0}
The semiclassical study of false vacuum decay in quantum field
theory has been performed by Callan and Coleman \cite{coleman}, in
close analogy with the work of Langer \cite{langer}. The
phenomenon occurs when the field theoretical potential
$U(\varphi)$ displays a relative minimum at $\varphi_{+}$: this
classical point corresponds to the false vacuum in the quantum
theory, which decays through tunnelling effects into the true
vacuum, associated with the absolute minimum $\varphi_{-}$ (see
Fig. \ref{figfalsevacpot}).

The main result of \cite{coleman} is the following expression for
the decay width per unit time and unit volume:
\begin{equation}
\frac{\Gamma}{V}\,=\,\left(\frac{B}{2\pi\hbar}\right)\,e^{-B/\hbar}
\left|\frac{\det'[-\partial^{2}+U''(\varphi)]}{\det[-\partial^{2}
+ U''(\varphi_{+})]}\right|^{-1/2}\,
\left[1+O(\hbar)\right]\;,
\end{equation}
specialized here to the case of two--dimensional space--time.
Omitting any discussion of the determinant, about which we refer
to the original papers \cite{coleman}, we will present here an
explicit analysis of the coefficient $B$.

It has been shown that $B$ coincides with the Euclidean action of
the so--called \lq\lq bounce" background $\varphi_{B}$:
\begin{equation}
\label{Bdecay}
B \,=\, S_{E}\,=\,2\pi\int\limits_{0}^{\infty}d\rho\,
\rho\left[\frac{1}{2}\left(\frac{d\varphi_{B}}{d\rho}\right)^{2}
+ U(\varphi_{B})\right]\;.
\end{equation}
This classical solution is the field--theoretical generalization
of the path of least resistance in quantum mechanical tunnelling;
it only depends on the Euclidean radius
$\rho\,=\,\sqrt{\tau^{2}+x^{2}}$ and satisfies the equation
\begin{equation}
\label{bounceeq}
\frac{d^{2}\varphi_{B}}{d\rho^{2}} +
\frac{1}{\rho}\,\frac{d\varphi_{B}}{d\rho}\,=\,U'[\varphi_{B}]\;,
\end{equation}
with boundary conditions
\begin{equation}
\lim\limits_{\rho\to\infty}
\varphi_{B}(\rho)\,=\,\varphi_{+}\;,\qquad
\frac{d\varphi_{B}}{d\rho}(0)=0\;.
\end{equation}

Although in general one does not know explicitly the bounce
solution, it is possible to set up some approximation to extract a
closed expression for the coefficient $B$. The so-called \lq\lq
thin wall" approximation consists in viewing the potential
$U(\varphi)$ as a perturbation of another potential
$U_{+}(\varphi)$, which displays degenerate vacua at
$\varphi_{\pm}$ and a kink $\varphi_{K}(x)$ interpolating between
them. The small parameter for the approximation is the energy
difference $\varepsilon=U(\varphi_{+})-U(\varphi_{-})$.

In this framework, one can qualitatively guess that the bounce has
a value $\varphi(0)$ very close to $\varphi_{-}$, then it remains
in this position until some vary large $\rho=R$ and finally it
moves quickly towards the final value $\varphi_{+}$. For $\rho$
near $R$, the first--derivative term in eq. (\ref{bounceeq}) can
be neglected; if in addition one also approximates $U$ with
$U_{+}$, then one can express the unknown bounce solution as
\cite{coleman}
\begin{equation}\label{colemanbounce}
\varphi_{B}(\rho)\,=\,
\begin{cases}
\;\varphi_{-}&\qquad \rho\ll R\\
\;\varphi_{K}(\rho-R)&\qquad \rho\approx R\\
\;\varphi_{+}&\qquad \rho\gg R\;.
\end{cases}
\end{equation}
Since the bounce has to represent the path of least resistance,
the parameter $R$, free up to this point, can be fixed by minimizing
the action
\begin{equation}
S_{E}\,=\,-\pi R^{2}\epsilon+2\pi R \,M_{K}\;,
\end{equation}
which is given by the sum of a volume term and a surface term.
Hence, the condition $\frac{dS_{E}}{dR}=0$ is realized by the
balance of these two different terms in competition, and it
finally gives
\begin{equation}\label{colemanpred}
R\,=\,\frac{M_{K}}{\varepsilon}\quad\Longrightarrow\quad
B\,=\,\pi\,\frac{M_{K}^{2}}{\varepsilon}\;.
\end{equation}

In the DSG model, however, we know explicitly the bounce
background in the thin wall regime (here we have
$\varepsilon=\frac{2\lambda}{\beta^{2}}$), without any
approximation on the potential. This is given by the solution
(\ref{bounce}) with $x$ replaced by $\rho$, that can be directly
used to estimate the decay width. Unfortunately the integral
in (\ref{Bdecay}) does not admit a simple expression to be
expanded for small $\lambda$, but it is clear from eq.\,(\ref{bouncesumSG})
and Fig. \ref{figbounce} that the leading contribution is given by
\begin{equation}
S_{E}\,\simeq\,2\pi R \int\limits_{R-\Delta r}^{R+\Delta
r}dx\left[\frac{d\varphi_{SG}}{dx}(R-x)\right]^{2}\,\simeq
\,\frac{8\pi}{\beta^{2}}\,\log\left(\frac{16\mu}{\lambda}\right)\;,
\end{equation}
with $R$ given by (\ref{fixedR}). This behavior in $\lambda$ does
not agree with the general prediction (\ref{colemanpred}). The
reason can be traced out in the fact that eq.\,(\ref{bouncesumSG})
explicitly realizes the relation between the bounce and the kink
of the unperturbed theory, but in a more sophisticated way than
(\ref{colemanbounce}). In fact, the mass parameter $m_{f}$ of the
SG kink $\varphi_{SG}$ is dressed to be the one of the DSG theory,
and the parameter $R$ is not free, since (\ref{bounce}) is already
the result of a minimization process, being a solution of the
Euler--Lagrange equations. The thin wall approximation can be
still consistently used because $R$ is very big for small
$\lambda$, while the crucial difference is that the volume term is
now missing from the action, since the value
$\varphi_{B}(0)=\varphi_{1}$ is the so--called classical turning
point (see Fig. \ref{figfalsevacpot}), degenerate with the false
vacuum. It is worth noting that the path of least resistance in
quantum mechanics precisely interpolates between the false vacuum
and the turning point.

Up to the determinant factor, our result for the leading term in
the decay width is then
\begin{equation}
\frac{\Gamma}{V}\,\simeq \,\frac{4}{\beta^{2}}\,
\left(\frac{\lambda}{16\mu}\right)^{8\pi/\beta^{2}}\,
\log\left(\frac{16\mu}{\lambda}\right)
\;.
\end{equation}
It will be interesting to investigate whether the above mentioned
difference with the prediction (\ref{colemanpred}) is a particular
feature of the DSG model or it appears for a generic potential if
one improves the approximate description of the bounce along the
lines discussed here.

\section{Other kind of resonances}\label{secres}
\setcounter{equation}{0}
The appearance of resonances in the classical scattering of the
Double Sine-Gordon kinks has been extensively studied with numerical
techniques, and a complete picture of this phenomenon can be found
in \cite{sodano}. In this work, the key ingredient for the
presence of resonances was identified in the presence of a discrete
eigenvalue, besides the zero mode, of the small oscillations around
the kink background. This eigenvalue, called \lq\lq shape mode",
represents an internal excitation of the kink \cite{DHN,raj}.

This mechanism can be easily interpreted also in our formalism,
but unfortunately in the case of the Double Sine-Gordon model we
were not able to solve analytically the stability equation around the
kink backgrounds. Hence, we will limit ourselves to the discussion
of the same phenomenon in a simpler theory, the $\phi^{4}$ field
theory in the broken symmetry phase\footnote{The main features of
the numerical analysis performed in \cite{sodano} for the DSG model
were indeed previously recognized in this simpler theory \cite{campbell}.}:
\begin{equation}\label{phi4pot}
V(\phi) \,=\, \frac{g}{4}\,\phi^{4}-\frac{m^{2}}{2}\,\phi^{2} +
\frac{m^{4}}{4 g}\;.
\end{equation}
The standard kink background of this theory is given by
\begin{equation}
\label{phi4kinkinf}
\phi_{cl}(x) \, = \,\frac{m}{\sqrt{g}}\, \tanh\left(
\frac{m x}{\sqrt{2}}\right)\;,
\end{equation}
with classical energy
$M\,=\,\frac{2\sqrt{2}}{3}\frac{m^{3}}{g}$. The small oscillations
(\ref{stability}) around this solution have, in addition to the usual
translational mode $\omega_{0}=0$, another discrete eigenvalue
\begin{equation}
\omega_{1}^{2}\,=\,\frac{3}{2}\,m^{2}\;,
\end{equation}
which represents an internal excitation of the kink \cite{DHN,raj}.
This feature, quite crucial for the analysis performed in \cite{campbell},
has a counterpart in our formalism. In fact, it was shown by Goldstone
and Jackiw \cite{goldstone} that, performing the Fourier transform of the
corresponding eigenfunction $\eta_{1}(a)$, one can write the form
factor of the field $\phi$ between asymptotic states containing a
simple and an excited kink. Furthermore, also this result, as the
previous one relative to the form factors of the elementary kinks,
can be refined in terms of the rapidity variable, so that one
obtains a covariant expression that can be analytically continued
in the crossed channel. Since in this case the eigenfunction is
\begin{equation}
\eta_{1}(x)\,=\,-\frac{\sinh\left(\frac{mx}{\sqrt{2}}\right)}
{2\cosh^{2}\left(\frac{mx}{\sqrt{2}}\right)} \; ,
\end{equation}
for the corresponding form factor we have
\begin{equation}
\langle\, 0|\,\phi(0)\,|\,\bar{p}_{2}\,p^{*}_{1}\,\rangle
\, = \,-i\,\frac{M\,\pi}{6^{1/4}\,m^{5/2}}\,
\frac{M\,(i\pi-\theta)}{\cosh\left[\frac{\pi}{\sqrt{2}\,m}\,
M\,(i\pi-\theta)\right]} \;,
\end{equation}
where the $p^{*}_1$ denotes the momentum of the excited kink state.
The dynamical poles of this object correspond to bound states with
masses
\begin{equation}
\label{phi4excboundst}
\left(m_{b^{*}}^{(n)}\right)^{2} \, = \,
4 M (M + \omega_{1}) \sin^{2}\left[\frac{3}{8}\,
\frac{g}{m^{2}}\,(2n+1)\right] + \omega_{1}^{2} \;.
\end{equation}
The states with
\begin{equation}
\label{resnumber}
\frac{8}{3}\,\frac{m^{2}}{g}\,
\arcsin\sqrt{\frac{4M^{2} - \omega_{1}^{2}}{4M ( M + \omega_{1})}}
< 2n+1 < \frac{4}{3}\,\frac{m^{2}}{g}\,\pi
\end{equation}
have masses in the range
\begin{equation}
2M <m_{b^{*}}^{(n)} < 2M + \omega_{1}\;,
\end{equation}
and, therefore, they can be seen as resonances in the kink-antikink
scattering.

Since the numerical analysis done in \cite{campbell} is independent of
the coupling constant\footnote{Classically, in fact, one can always
rescale the field and eliminate the coupling constant $g$.}, a
quantitative comparison with our semiclassical result is rather difficult,
due to the dependence on $g$ of (\ref{resnumber}). However, the presence
of many resonance states seen at classical level is qualitatively
confirmed to persist also in the quantum field theory at small $g$,
i.e. in its semiclassical regime, according to (\ref{resnumber}).

Back to the Double Sine-Gordon model, the shape mode with the
relative resonances has been numerically observed for the small
kink (\ref{smallkink}) in the $\delta = \frac{\pi}{2}$ case, and
for the kink (\ref{doublekink}) in the case $\delta = 0$. Our
analysis is in agreement with these results, and it adds another
possibility for the small kink case. In fact, since in this regime
it is also present the large kink, which has higher mass, the
resonances seen in the small kink-antikink scattering are related
both to their excited bound states with masses $m_{S^{*}}^{(n)}$
in the range
\begin{equation}
2M_{S} < m_{S^{*}}^{(n)} < 2M_{S} + \tilde{\omega}_{1}\;,
\end{equation}
and to the large kink-antikink bound states with masses in the
range
\begin{equation}
2M_{S} < m_{L}^{(n)} < 2M_{L}\;,
\end{equation}
where $ m_{L}^{(n)}$ are given by (\ref{largebound}).

\section{Conclusions}\label{secconcl}
\setcounter{equation}{0}
When available, the Semiclassical Method is an efficient tool for
studying the mass spectrum of an integrable or a non--integrable
theory. In the last case, it may be complementary to the Form
Factor Perturbation Theory or it may provide results comparable
with this method. We have applied both techniques for analysing
the mass spectrum of the non--integrable quantum field theory
given by the Double Sine--Gordon model, for few qualitatively
different regions of its coupling--constants space. This model
appears to be an ideal theoretical playground for understanding
some of the relevant features of non--integrable models. By moving
its coupling constants, it shows, in fact, different types of kink
excitations and confinement phenomena, a rich spectrum of meson
particles, resonance states, false vacuum decay and the occurrence
of a phase transition. In light of the many applications it finds
in condensed matter systems, it would be interesting to
investigate further its properties.

\vspace{1cm}

\begin{flushleft}\large
\textbf{Acknowledgements}
\end{flushleft}
The authors thank G. Delfino for valuable discussions. Two of us
(G.M. and V.R.) would also like to thank the Instituto de Fisica Teorica
in San Paulo for the warm hospitality during the period of their staying,
when this work was done. We also ackowledge interesting discussions with
Z. Bajnok, L. Palla, G. Takacs and F. Wagner. Moreover, G.M. thanks them
for the nice hospitality during his visit to Budapest. This work was partially
supported by the Italian COFIN contract ``Teoria dei Campi, Meccanica
Statistica e Sistemi Elettronici'' and by the European Commission TMR
programme HPRN-CT-2002-00325 (EUCLID). G.S. thanks FAPESP for the finantial
support.

\newpage

\begin{appendix}

\section{Kink mass corrections in the FFPT}\label{kmasscorr}
\setcounter{equation}{0}
In this Appendix we compute by means of the FFPT the corrections
to the kink masses in the semiclassical limit, which is relevant
for a comparison with our results.

For small $\lambda$, we have to consider the DSG model as a
perturbation of the two--folded Sine-Gordon \cite{tak}. In the
$\delta = \pi/2$ case, the perturbing operator is $\Psi =
\sin\frac{\beta}{2}\phi$. Its form factors between the
vacuum and the two possible kink-antikink asymptotic states are
obtained at the semiclassical level by performing the Fourier transform
of $\;\sin\left[\frac{\beta}{2}K_{k,k+1}^{cl}(x)\right]$
\cite{MRS}, with $K_{k,k+1}^{cl}(x)$ given by eq.\,(\ref{twofoldedkinks}).
Hence we obtain
\begin{equation}
F_{K_{k,k+1},\bar{K}_{k,k+1}}^{\Psi}(\theta)\, =\,
\frac{8\pi}{\beta^{2}}\,(-1)^{k}\,\frac{1}
{\cosh\frac{4\pi}{\beta^{2}}(\theta-i\pi)}\;.
\end{equation}
The first order correction in $\lambda$ to the kink masses is then
\begin{equation}
\delta
M_{K_{k,k+1}}\,=\,
\frac{\lambda}{\beta^{2}}\,\frac{1}{M_{K}}
\,F_{K_{k,k+1},\bar{K}_{k,k+1}}^{\Psi}(i\pi)
\,=\,(-1)^{k}\frac{\lambda}{\beta^{2}}\,\frac{\pi}{\sqrt{\mu}}\;,
\end{equation}
in agreement with the correction to the classical masses
(\ref{largesmallkinkfirst}), since $K_{0,1}$ is associated with
the large kink, and $K_{1,2}$ with the small one.

In the $\delta = 0$ case, instead, we can explicitly see how the
solitons disappear from the spectrum as soon as $\lambda$ is switched
on. The form factor of the operator $\Psi = \cos\frac{\beta}{2}\phi$
has, in fact, a divergence at $\theta=i\pi$
\begin{equation}
F_{K_{k,k+1},\bar{K}_{k,k+1}}^{\Psi}(\theta)\, =\,-i
\frac{8\pi}{\beta^{2}}\,(-1)^{k}\,\frac{1}
{\sinh\frac{4\pi}{\beta^{2}}(i\pi-\theta)}\;.
\end{equation}

\vspace{0.5cm}

The other interesting regime to explore is the small $\mu$ limit.
In the case $\delta=0$, this can be seen as the perturbation of
the SG model at coupling $\tilde{\beta} = \beta/2$ by means of the
operator $\Psi = \cos\,2\tilde{\beta}\varphi$. The semiclassical form
factor is
\begin{equation}
F_{K,\bar{K}}^{\Psi}(\theta)\, =\,
\frac{16}{3}\,\frac{32}{\beta^{2}}\,\frac{i\pi y}
{\sin\,i\pi y}(1-2\,y^{2})\;,
\end{equation}
where we have defined $y =\frac{16}{\beta^{2}}(i\pi-\theta)$. The
corresponding mass correction is given by
\begin{equation}
\delta
M_{K}\,=\,\frac{\mu}{\beta^{2}}\,\frac{1}{M_{K}}\,
F_{K,\bar{K}}^{\Psi}(i\pi)\,=\,\frac{\mu}{\beta^{2}}\,
\frac{16}{3}\,\frac{1}{\sqrt{\lambda/4}}\;,
\end{equation}
in agreement with (\ref{doublekinkfirst}).

The case $\delta = \frac{\pi}{2}$ can be described by shifting the
original SG field as $\varphi\, \to\,\varphi + \frac{\pi}{\beta}$.
In this way the perturbing operator becomes $-\Psi$ and we finally
obtain the same mass correction but with opposite sign, as in
(\ref{commonkinkmass}).

\section{Semiclassical form factors}\label{secFF}
\setcounter{equation}{0}
In this Appendix we explicitly present the expressions of the
two--particle form factors, on the asymptotic states given by
the different kinks appearing in the DSG theory, of the operators
$\varphi(x)$ and $\varepsilon(x)$, the last one defined by
$$
\varepsilon(x) \,\equiv \, \frac{1}{2}
\left(\frac{d\varphi}{dx}\right)^{2}+V[\varphi(x)] \,\,\,.
$$
These matrix elements are obtained by performing the Fourier
transforms of the corresponding classical backgrounds, as
indicated in (\ref{ffinf}) and (\ref{f2}). We use the notation:
$$
F^{\Psi}_{K\bar{K}}(\theta)\,=\,\langle\,
0\,|\,\Psi(0)\,|\,K(\theta_1)\,\bar{K}(\theta_2)\,\rangle\;,
$$
with $\theta = \theta_1 - \theta_2$.
\vspace{0.5cm}

For the kink (\ref{doublekink}) in the $\delta=0$ case we have
\begin{equation}
F^{\varphi}_{K\bar{K}}(\theta)\,=\,\frac{4\pi^{2}}{\beta}\,M_{K}\,
\delta\left[M_{K}(i\pi-\theta)\right] + i\,\frac{4\pi}{\beta}\,
\frac{1}{i\pi-\theta}\;\frac{\cos\left[\alpha
\,\frac{M_{K}}{m}\,(i\pi-\theta)\right]}{\cosh\left[\frac{\pi}{2}
\,\frac{M_{K}}{m}\,(i\pi-\theta)\right]}\;,
\end{equation}
where
$$
\alpha\,=\,\text{arccosh}\sqrt{\frac{\lambda+4\mu}{\lambda}}\;,
$$
while $m$ and $M_{K}$ are given by (\ref{curvaturedeltazero}) and
(\ref{doublekinkmass}), respectively, and
\begin{equation}
F^{\varepsilon}_{K\bar{K}}(\theta)\,=\,-\,\frac{128\pi}{\beta^{2}}\,
\frac{m^{3}
M_{K}}{\lambda}\;\left\{\frac{1}{\sinh\left[\pi\,\frac{M_{K}}{2m}\,
(i\pi-\theta)\right]}\;\frac{d}{dc} \left[\frac{\sinh\left[(\text{arccosh}\, c)
\frac{M_{K}}{2m}(i\pi-\theta)\right]}{\sqrt{c^{2}-1}}\right]+\right.
\end{equation}
\begin{equation*}
\left.
-\frac{2\sinh\pi}{\cosh\left[\pi\,\frac{M_{K}}{m}\,(i\pi-\theta)\right]-1}
\;\frac{d}{dc} \left[\frac{c\,\sinh\left[(\text{arccosh} \,c)
\frac{M_{K}}{2m}(i\pi-\theta)\right]}{\sqrt{c^{2}-1}}\right]\right\}\;,
\end{equation*}
where $c=1+\frac{8\mu}{\lambda}$.

\vspace{0.5cm}

For the large kink (\ref{largekink}) in the $\delta=\frac{\pi}{2}$
case (with $\lambda<4\mu$) we have
\begin{equation}
F^{\varphi}_{L\bar{L}}(\theta)\,=\,\frac{2\pi^{2}}{\beta}\,M_{L}\,
\delta\left[M_{L}(i\pi-\theta)\right] +
i\,\frac{4\pi}{\beta}\,
\frac{1}{i\pi-\theta}\;\frac{\sinh\left[\alpha
\,\frac{M_{L}}{m}\,(i\pi-\theta)\right]}{\sinh\left[\pi\,
\frac{M_{L}}{m}\,(i\pi-\theta)\right]}\;,
\end{equation}
where
$$
\alpha\,=\,2\arctan\sqrt{\frac{4\mu+\lambda}{4\mu-\lambda}}\;,
$$
while $m$ and $M_{L}$ are given by (\ref{curvaturedeltapi2}) and
(\ref{largesmallkinkmass}), respectively, and
\begin{equation}
F^{\varepsilon}_{L\bar{L}}(\theta)\,=\,\frac{8\pi}{\beta^{2}}\,
\frac{m^{3}
M_{L}}{\mu}\;\frac{1}{\sinh\left[\pi\,\frac{M_{L}}{m}\,(i\pi-\theta)\right]}\;
\frac{d}{dc} \left\{\frac{\sinh\left[(\arccos\, c)
\frac{M_{L}}{m}(i\pi-\theta)\right]}{\sqrt{1-c^{2}}}\right\}\;,
\end{equation}
where $c=-\frac{\lambda}{4\mu}$.

\vspace{0.5cm}

For the small kink (\ref{smallkink}) in the $\delta=\frac{\pi}{2}$
case (with $\lambda<4\mu$) we have
\begin{equation}
F^{\varphi}_{S\bar{S}}(\theta)\,=\,\frac{6\pi^{2}}{\beta}\,M_{S}\,
\delta\left[M_{S}(i\pi-\theta)\right] +
i\,\frac{4\pi}{\beta}\,
\frac{1}{i\pi-\theta}\;\frac{\sinh\left[\alpha
\,\frac{M_{S}}{m}\,(i\pi-\theta)\right]}{\sinh\left[\pi\,
\frac{M_{S}}{m}\,(i\pi-\theta)\right]}\;,
\end{equation}
where
$$
\alpha\,=\,2\arctan\sqrt{\frac{4\mu-\lambda}{4\mu+\lambda}}\;,
$$
while $m$ and $M_{S}$ are given by (\ref{curvaturedeltapi2}) and
(\ref{largesmallkinkmass}), respectively, and
\begin{equation}
F^{\varepsilon}_{S\bar{S}}(\theta)\,=\,\frac{8\pi}{\beta^{2}}\,
\frac{m^{3}
M_{S}}{\mu}\;\frac{1}{\sinh\left[\pi\,\frac{M_{S}}{m}\,(i\pi-\theta)\right]}\;
\frac{d}{dc} \left\{\frac{\sinh\left[(\arccos \, c)
\frac{M_{S}}{m}(i\pi-\theta)\right]}{\sqrt{1-c^{2}}}\right\}\;,
\end{equation}
where $c=\frac{\lambda}{4\mu}$.

\vspace{0.5cm}

Finally, for the kink (\ref{commonkink}) in the
$\delta = \frac{\pi}{2}$ case (with $\lambda>4\mu$) we have
\begin{equation}
F^{\varphi}_{K\bar{K}}(\theta)\,=\,\frac{2\pi^{2}}{\beta}\,M_{K}\,
\delta\left[M_{K}(i\pi-\theta)\right] +
i\,\frac{4\pi}{\beta}\,
\frac{1}{i\pi-\theta}\;\frac{\cos\left[\alpha
\,\frac{M_{K}}{m}\,(i\pi-\theta)\right]}
{\cosh\left[\frac{\pi}{2}\,\frac{M_{K}}{m}\,(i\pi-\theta)\right]}\;,
\end{equation}
where
$$
\alpha\,=\,\text{arccosh}\sqrt{\frac{\lambda-4\mu}{\lambda}}\;,
$$
while $m$ and $M_{L}$ are given by (\ref{curvaturedeltapi2sec})
and (\ref{commonkinkmass}), respectively, and
\begin{equation}
F^{\varepsilon}_{K\bar{K}}(\theta)\,=\,-\,\frac{128\pi}{\beta^{2}}\,
\frac{m^{3}
M_{K}}{\lambda}\;\left\{\frac{1}{\sinh\left[\pi\,\frac{M_{K}}{2m}\,
(i\pi-\theta)\right]}\;\frac{d}{dc} \left[\frac{\sinh\left[(\arccos \, c)
\frac{M_{K}}{2m}(i\pi-\theta)\right]}{\sqrt{1-c^{2}}}\right]+\right.
\end{equation}
\begin{equation*}
\left.
-\frac{2\sinh\pi}{\cosh\left[\pi\,\frac{M_{K}}{m}\,(i\pi-\theta)\right]-1}\;
\frac{d}{dc} \left[\frac{c\,\sinh\left[(\arccos\, c)
\frac{M_{K}}{2m}(i\pi-\theta)\right]}{\sqrt{1-c^{2}}}\right]\right\}\;,
\end{equation*}
where $c=1-\frac{8\mu}{\lambda}$.

\section{Neutral states in the $\delta=\frac{\pi}{2}$ case}\label{breathers}
\setcounter{equation}{0}
The semiclassical results reported in the text, i.e.
eqs.\,(\ref{largebound}), (\ref{smallbound}) and
(\ref{linearlambda}), pose an interesting question about the
nature of neutral states in the DSG model at $\delta=\frac{\pi}{2}$.
It should be noticed, in fact, that the first order correction in
$\lambda$ obtained by the Semiclassical Method does not match with
the results reported in \cite{tak} where, by using the FFPT and an
extrapolation of numerical data, the authors concluded that this
correction was instead identically zero\footnote{It is worth stressing
that the linear correction (\ref{linearlambda}) in $\lambda$ is very
small even for finite values of $\beta$ (it is easy to check, indeed,
that the first term of its expansion is $\frac{\pi}{24}
\left(\frac{\beta^2}{16}\right)^2$) and somehow compatible with the numerical data
given in \cite{tak}.}. It is worth discussing this problem
in more detail.

In the standard Sine--Gordon model, the breathers $|\,b_{n}\rangle$,
with $n$ odd (or even), are defined as the bound states of odd (or
even) combinations of $|\,K\,\bar{K}\rangle$ and $|\,\bar{K}\,K\rangle$,
where $K$ represents the soliton and $\bar{K}$ the antisoliton. The
combinations $|\,K\,\bar{K}\pm\bar{K}\,K\rangle$ are eigenstates
of the parity operator $P:\; \phi\to - \phi$, which commutes with
the hamiltonian and acts on the soliton transforming it into the
antisoliton. The above mentioned identification of the bound
states relies on a very peculiar feature of the Sine--Gordon
$S$--matrix in the soliton sector \cite{zams}, whose elements are
defined as
\begin{eqnarray}
K(\theta_{1})\,\bar{K}(\theta_{2})&=&
S_{T}(\theta_{12})\,\bar{K}(\theta_{2})\,K(\theta_{1})+
S_{R}(\theta_{12})\,K(\theta_{2})\,\bar{K}(\theta_{1})\;,\\
K(\theta_{1})\,K(\theta_{2})&=&
S(\theta_{12})\,K(\theta_{2})\,K(\theta_{1})\;,\\
\bar{K}(\theta_{1})\,\bar{K}(\theta_{2})&=&
S(\theta_{12})\,\bar{K}(\theta_{2})\,\bar{K}(\theta_{1})\;.
\end{eqnarray}
In fact, both the transmission and the reflection amplitudes
$S_{T}(\theta)$ and $S_{R}(\theta)$ display poles at $\theta_{n}^{*} =
i(\pi-n\xi)$, with residua which are equal or opposite in sign
depending whether $n$ is odd or even. Hence, the diagonal elements
\begin{eqnarray}
S_{-}(\theta)&=&
\frac{1}{2}\left[S_{T}(\theta)-S_{R}(\theta)\right]\;,\\
S_{+}(\theta)&=&
\frac{1}{2}\left[S_{T}(\theta)+S_{R}(\theta)\right]
\end{eqnarray}
have only the poles with odd or even $n$, respectively, and for
each $n$ there is only one bound state with definite parity.

However, this is a special feature of the Sine--Gordon model which
finds no counterpart, for instance, in other problems with a similar
structure. As an explicit example, one can consider the $(\text{RSOS})_{3}$
scattering theory, which displays a 3-fold degenerate vacuum and
two types of kink and antikink with the same mass. The central
vacuum is surrounded by two other minima, as in the Sine--Gordon
case, and this gives the possibility to define both a
kink-antikink state and an antikink-kink state around it. The
minimal scattering matrix, given in \cite{rsos}, can be dressed
with a CDD factor to generate bound states. It is easy to check
that the common poles in the transmission and reflection amplitudes
have in this case different residua, giving rise to two distinct
bound states, degenerate in mass, over the central vacuum.

Hence, if we call $|\,b_{n}^{(0)}\rangle$ the bound states of
kink-antikink and $|\,b_{n}^{(1)}\rangle$ the bound states of
antikink-kink, in general we have to consider them as two distinct
excitations, and if they have the same mass we can build two other
states from their linear combinations
\begin{equation}\label{basischange}
|\,b_{n}^{(\pm)}\rangle\,=\,\frac{|\,b_{n}^{(0)}\rangle\,\pm\,
|\,b_{n}^{(1)}\rangle}{\sqrt{2}}\;.
\end{equation}
The peculiarity of the Sine--Gordon model is the removal of this
double multiplicity due to the fact that the states
$|\,b_{2n+1}^{(+)}\rangle$ and $|\,b_{2n}^{(-)}\rangle$ decouple
from the theory. This feature is shared also by the two-folded
version of the model, since the kink scattering amplitudes have
the same analytical form as in SG \cite{foldedSG}.

In the two--folded SG there are two different kink states
$|\,K_{-1,0} \rangle$ and $|\,K_{0,1}\rangle$ (see Sect.
\ref{secDSG} and ref. \cite{foldedSG} for the notation), and the
parity $P$, which is still an exact symmetry of the theory, acts
on them transforming the kink of one type into the antikink of the
other type:
\begin{equation}
P: \;|\, K_{0,1}\rangle\to\,|\,K_{0,-1}\rangle\;,\quad
|\,K_{-1,0}\rangle\to\,|\,K_{1,0}\rangle\;.
\end{equation}
If we quantize the theory around the vacuum $|\,0\rangle$, we can
define $|\,b_{n}^{(0)} \rangle$ as the bound states of
$|\,K_{0,1}\,K_{1,0}\rangle$, and $|\,b_{n}^{(1)}\rangle$ as the
bound states of $|\,K_{0,-1}\,K_{-1,0}\rangle$. These degenerate
states, which transform under $P$ as
\begin{equation}
P: \; |\,b_{n}^{(0)}\rangle\to\,|\,b_{n}^{(1)}\rangle\;,\quad
|\,b_{n}^{(1)}\rangle\to\,|\,b_{n}^{(0)}\rangle\;,
\end{equation}
can be still organized in parity eigenstates
$|\,b_{n}^{(\pm)}\rangle$, and the particular dynamics of the
problem causes the decoupling of half of them from the theory.
Furthermore, it is easy to see that the form factors of an odd
operator between two of these states has to vanish in virtue of
the relation
\begin{eqnarray*}
\langle\,0\,|\,\sin\frac{\beta}{2}\phi\,|\,b_{n}^{(\pm)}b_{n}^{(\pm)}\rangle
& = &
\langle\,0\,|\,P^{-1}P\,\left(\sin\frac{\beta}{2}\phi\right)\,
P^{-1}P|\,b_{n}^{(\pm)}b_{n}^{(\pm)}\rangle \,=\\
& = &-\,
\langle\,0\,|\,\sin\frac{\beta}{2}\phi\,|\,b_{n}^{(\pm)}b_{n}^{(\pm)}\rangle\;,
\end{eqnarray*}
leading to the FFPT result that the breathers receive a zero mass
correction at first order in $\lambda$, as it is claimed in
\cite{tak}.

However, FFPT can be applied by taking into account the nature of
neutral states in the DSG model, where the addition to the
Lagrangian of the term
$\;-\frac{\lambda}{\beta^{2}}\,\sin\frac{\beta}{2}\varphi\;$
spoils the invariance under $P$. The kinks $|\,K_{-1,0} \rangle$
and $|\,K_{0,1} \rangle$ are deformed into the small and large
kinks $|\,S \rangle$ and $|\,L \rangle$, respectively, which are
not anymore degenerate in mass and cannot be superposed in linear
combinations. Hence, the neutral states present in the theory are
$|\,b_{n}^{(L)}\rangle$ and $|\,b_{n}^{(S)}\rangle$, deformations
of  $|\,b_{n}^{(0)}\rangle$ and $|\,b_{n}^{(1)}\rangle$
respectively. In virtue of the general considerations presented
above, one can see that this interpretation does not lead to any
drastic change in the spectrum. In fact, the states
$|\,b_{2n+1}^{(+)}\rangle$ and $|\,b_{2n}^{(-)}\rangle$ have no
reason to decouple in the DSG theory, but they have to carry a
coupling which is a function of $\lambda$ adiabatically going to
zero in the two--folded SG limit.

A proper use of the FFPT on $|\,b_{n}^{(0)}\rangle$ and
$|\,b_{n}^{(1)}\rangle$ reproduces indeed the situation described
by (\ref{linearlambda}), in which the two sets of breathers
receive mass corrections including also odd terms in $\lambda$,
but with opposite signs. This is easily seen by considering the
$P$ transformations in the two--folded SG model:
\begin{eqnarray*}
\langle\,0\,|\,\sin\frac{\beta}{2}\phi\,|\,b_{n}^{(0)}b_{n}^{(0)}\rangle
& = &
\langle\,0\,|\,P^{-1}P\,\left(\sin\frac{\beta}{2}\phi\right)\,
P^{-1}P|\,b_{n}^{(0)}b_{n}^{(0)}\rangle \,=\\
& = &-\,
\langle\,0\,|\,\sin\frac{\beta}{2}\phi\,|\,b_{n}^{(1)}b_{n}^{(1)}\rangle\;,
\end{eqnarray*}
which gives, at first order in $\lambda$,
\begin{equation}
\delta\,m_{(L)}=-\,\delta\,m_{(S)}\;,
\end{equation}
in agreement with our semiclassical result (\ref{linearlambda}).
It is worth noting that also with this interpretation the total
spectrum of the DSG model remains unchanged under the action of
$P$, which corresponds to the transformation $\lambda\to
-\lambda$. In fact, the two types of kinks and breathers are
mapped one into the other. This is consistent with the observation
that $P$, although it is not anymore a symmetry of the perturbed
theory, simply realizes a reflection of the potential, hence the
total spectrum should be invariant under it.

Presently the above symmetry considerations seem to us the correct
criterion to define the neutral states, and find confirmation in our
semiclassical result (\ref{linearlambda}). However, the available
numerical data presented in \cite{tak} pose a challenge to this
interpretation and further studies are needed to solve this interesting
and delicate problem. In fact, although $\delta\,m_{(L)}$ and $\delta\,m_{(S)}$
are not forced to vanish by symmetry arguments, there is in principle the
possibility that both of them are identically zero in the complete
quantum computation. This could follow from the use of the exact
kink masses entering eqs.\,(\ref{largebound}) and (\ref{smallbound}),
together with a proper shift of the semiclassical pole in the form
factors, due to higher order contributions. The exact cancellation
of the linear corrections is a very strong requirement, in support
of which we have presently no indication in the theory, but a careful
analysis of this point is neverthless an interesting open
problem.

\section{Double Sinh--Gordon model}\label{secDShG}
\setcounter{equation}{0}

Among the different qualitative features taking place in perturbing
integrable models, a situation particularly interesting is the one
in which the perturbation is adiabatic for small values of the parameters
but nevertheless a qualitative changes in the spectrum occurs by increasing
its intensity.

This is indeed the situation in the $\delta = \frac{\pi}{2}$ case
of DSG model, where we have two types of kinks for small $\lambda$,
but at $\lambda = 4 \mu$ one of them disappears from the spectrum.
This phenomenon is obviously unaccessible by means of FFPT, hence
the semiclassical method is the best tool to describe it.

Here we consider another interesting example of this kind, realized
by the Double Sinh-Gordon Model (DShG). In this case the phenomenon
is even more evident, because in the unperturbed Sinh-Gordon model
there are no kinks at all, but just one scalar particle, while perturbing
it, at some critical value of the coupling a kink and antikink appear,
i.e. there is a deconfinement phase transition of these particles.

\begin{figure}[h]
\begin{tabular}{p{5cm}p{5cm}p{5cm}}
\psfrag{V}{$V(\mu=1,\,\lambda=0)$}
\psfig{figure=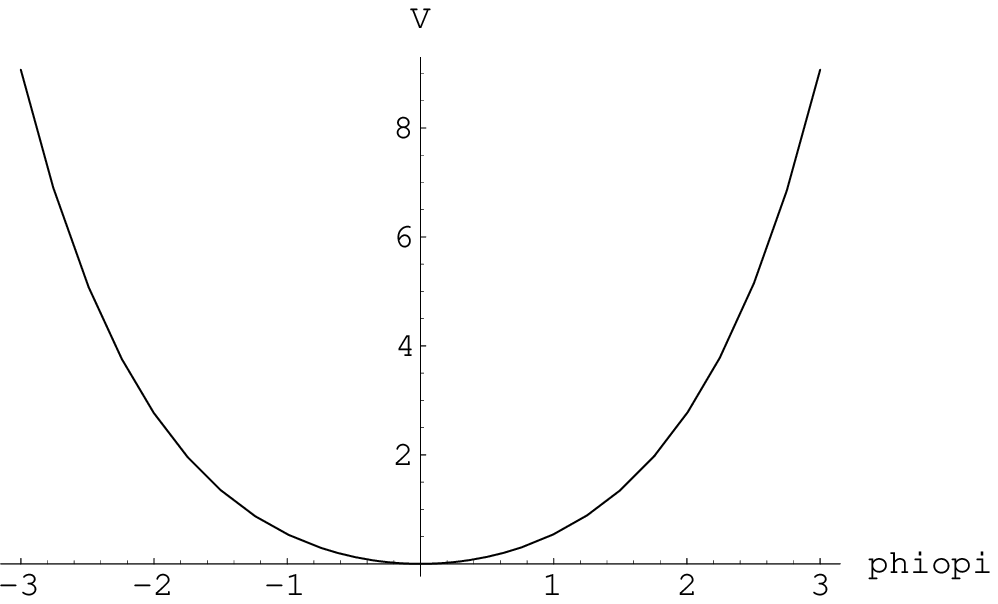,height=4cm,width=4.5cm}&
\psfrag{V}{$V(\mu=1,\,\lambda=4)$}\psfig{figure=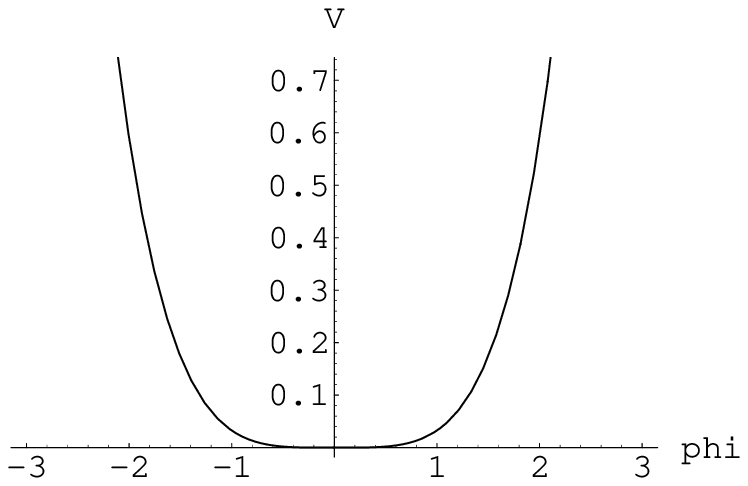,
height=4cm,width=4.5cm}& \psfrag{V}{$V(\mu=1,\,\lambda=5)$}
\psfig{figure=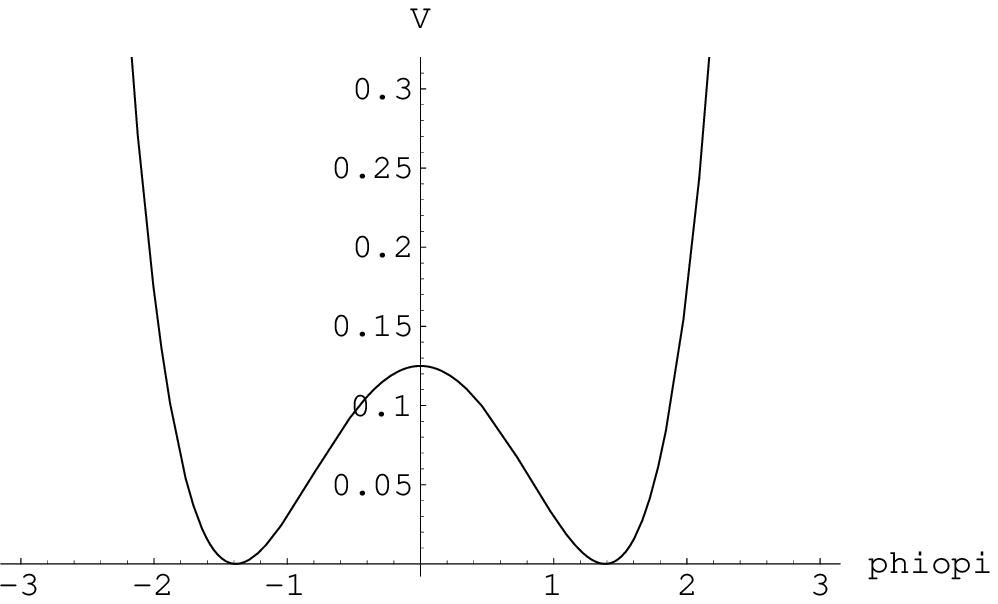,height=4cm,width=4.5cm}
\end{tabular}
\caption{DShG potential}\label{figpotDShG}
\end{figure}

The DShG potential, shown in Fig. \ref{figpotDShG}, is expressed
as
\begin{equation}\label{DShGpot}
V(\varphi) \,= \, \frac{\mu}{\beta^{2}}\,
\cosh\beta\,\varphi - \frac{\lambda}{\beta^{2}}\,
\cosh\left(\frac{\beta}{2}\,\varphi \right)\;.
\end{equation}

\psfrag{phiopi}{$\phi$}

In the regime $\lambda < 4 \mu$ the qualitative features are the
same as in the unperturbed Sinh-Gordon model. At $\lambda = 4
\mu$, however, the single minimum splits in two degenerate minima,
which for $\lambda > 4 \mu$ are located at $\varphi_{\pm} = \pm
\frac{2}{\beta} \text{arccosh}\frac{\lambda}{4\mu}$. A study of
the classical thermodynamical properties of the theory in this
regime has been performed in \cite{DShG} with the transfer
integral method.

The kink interpolating between the two degenerate vacua is
\begin{equation}
\varphi_{K}(x) \, = \, \frac{4}{\beta}\,\text{arctanh}
\left[\sqrt{\frac{\lambda-4\mu}{\lambda+4\mu}}\,
\tanh\left(\frac{m}{2}\,x\right)\right] \; ,
\end{equation}
with
$$
m^{2} \,= \,\frac{\lambda^{2}-16\mu^{2}}{16\mu}
\;.
$$
Its classical mass is given by
\begin{equation}
M_{K} \,=\, \frac{8m}{\beta^{2}}\left\{ -1 +
\frac{2\lambda}{\sqrt{\lambda^{2} - 16\mu^{2}}}\,\text{arctanh}
\sqrt{\frac{\lambda - 4\mu}{\lambda + 4\mu}}\right\}\;.
\end{equation}
From the form factor of $\varphi$ on the kink-antikink asymptotic
state, expressed as
\begin{equation}
F_{2}(\theta)\, =\, -i\frac{\pi}{\beta}\,
\frac{1}{i\pi-\theta}\,\frac{\sin\left[\text{arccosh}
\frac{\lambda}{4\mu}\, \frac{M_{K}}{m}\,(i\pi-\theta)\right]}
{\sinh\left[\pi\,\frac{M_{K}}{m}\,(i\pi - \theta)\right]}\;,
\end{equation}
we derive the bound states spectrum
\begin{equation}
m_{(K)}^{(n)}\,=\, 2 M_{K} \sin\left(n\,\frac{m}{2M_{K}}\right)
\,\,\,\,\,\,\,
,
\,\,\,\,\,\,\,
0 < n < \pi\frac{M_{K}}{m}
\end{equation}
All the kink--antikink bound states disappear from the theory at a
certain value $\lambda^{*}>4\mu$ such that
$\left.\pi\frac{M_{K}}{m}\right|_{\lambda^{*}}=1$, and the kink
becomes a constant solution with zero classical energy when
$\lambda \to 4 \mu$. This is the semiclassical manifestation of a
phase transition, analogous to the one observed in DSG with
$\delta=\frac{\pi}{2}$. As we have already anticipated, here the
phenomenon occurs in a reverse order, since in this case a kink
appears in the theory by increasing the coupling $\lambda$.

\end{appendix}

\end{document}